\documentclass[twocolumn,eqsecnum,aps,prc,tightenlines,floats]{revtex4}

\def\bea{\begin{eqnarray}}
\def\eea{\end{eqnarray}}

\newcommand{\bra}[1]{\left\langle #1\right|}
\newcommand{\ket}[1]{\left|#1\right\rangle}
\newcommand{\braket}[2]{\langle #1|#2\rangle}

\def\be{\begin{equation}}
\def\ee{\end{equation}}
\def\ba{\begin{eqnarray}}
\def\ea{\end{eqnarray}}

\usepackage{graphics}
\usepackage{graphicx}
\usepackage{epsf} 
\usepackage{slashed}
\usepackage{amsmath}
\usepackage{amssymb}
\def\sfrac#1#2{{\textstyle \frac{#1}{#2}}}

\begin{document} 


\phantom{0}
\vspace{-0.2in}
\hspace{5.5in}
\parbox{1.5in}{ \leftline{JLAB-THY-08-804}} 

\vspace{-1in}
\title
{\bf A Covariant model for the nucleon and the $\Delta$}
\author{G. Ramalho$^{1,2}$,  M.T. Pe\~na$^{2,3}$ and Franz Gross$^{1,4}$
\vspace{-0.1in}  }

\affiliation{
$^1$Thomas Jefferson National Accelerator Facility, Newport News, 
VA 23606 \vspace{-0.15in} }
\affiliation{
$^2$Centro de F{\'\i}sica Te\'orica de Part\' \i culas, 
Av.\ Rovisco Pais, 1049-001 Lisboa, Portugal \vspace{-0.15in}}
\affiliation{
$^3$Department of Physics, Instituto Superior T\'ecnico, 
Av.\ Rovisco Pais, 1049-001 Lisboa, Portugal \vspace{-0.15in} }
\affiliation{
$^4$College of William and Mary, Williamsburg, Virginia 23185}

\date{\today}
\begin{abstract}

The covariant spectator formalism is used to model the nucleon and the $\Delta$(1232)
as a system of three constituent quarks with their own  electromagnetic structure.
The definition of the ``fixed-axis'' polarization states for the diquark
emitted from the initial state vertex and absorbed into the final state
vertex is discussed. The helicity sum over those states is evaluated and seen to be covariant.
Using this approach, all four 
electromagnetic form factors of the nucleon, together with
the {\it magnetic\/} form factor, $G_M^*$, for the 
$\gamma N \rightarrow \Delta$ transition,
can be described using manifestly covariant 
nucleon and $\Delta$ wave functions with {\it zero\/}
orbital angular momentum $L$, but a successful description of $G_M^*$ near $Q^2=0$ requires the addition of a pion cloud term not included in the class of valence quark models considered here.  
We also show that the pure $S$-wave model gives electric, $G_E^*$, and
coulomb, $G^*_C$, transition form factors 
that are identically zero, showing that these 
form factors are sensitive to wave function components with $L>0$. 

\end{abstract}
\phantom{0}
\vspace*{0.9in}  
\maketitle

\section{Introduction}

Experimentally the main source of information
on  the internal structure of baryons lies in
the electro- and photo-excitation of the nucleon, and is
parametrized in terms
of electromagnetic form factors.
The precise elastic electron-proton polarization transfer
experiments undertaken at
Jlab \cite{Jones99,Gayou01,Punjabi05}  disclosed  results at
odds with previous 
data,  triggering 
an intense discussion 
about the shape of the nucleon \ \cite{Miller06,nucleon}   
(a review of the subject can be found 
in Refs.\ \cite{Hyde-Wright04,Arrington06}).  
Recent measurements of the $\gamma N\to\Delta$ 
transition raise further questions.
How much of this process is due to the three 
valence quarks in the $\Delta$, and how much
to the meson-nucleon, or ``pion cloud'' contribution?   
Experimental progress has been impressive, 
with the recent accumulation 
of high precision data over a wide momentum transfer 
($Q^2$) range. 
Just ten years ago, for example, the sign of the electric
and magnetic ratio for the $\gamma N\to\Delta$ transition,
was not even known beyond the photon point ($Q^2=0$).
New precise data  have been collected 
from MAMI \cite{MAMI},
LEGS \cite{LEGS}, MIT-Bates \cite{Bates}
and Jlab \cite{CLAS02,CLAS06}
in a region $Q^2 \le 6$ GeV$^2$
($q^2=-Q^2$ is the squared transferred momentum).

This is the second in a series of papers using the covariant spectator theory to study the implications of modeling the baryon resonances as covariant bound states of three valence constituent  quarks (CQ).  In this model the structure of the CQ (including an anomalous magnetic moment) arises from the dressed $\gamma\to q\bar q$ interactions which give rise to the familiar meson structure of the vector dominance model, and these contributions are {\it not\/} included in the wave function of the nucleon.  This language differs significantly from the light-cone formalism, where all of this vector dominance physics must be included in higher, non-valence components of the Fock-space expansion of the nucleon wave function (for further discussion, see \cite{nucleon}).     In our first paper \cite{nucleon} (referred to as Ref.\ I) we showed that a very simple pure $S$-wave model of the nucleon, based on a covariant generalization of a simple non-relativistic 
$SU(2)\times SU(2)$ wave function for three valence CQ, could describe the four elastic nucleon form-factors very well.  Our best model used 8 parameters: two to model the ``radial'' structure of the nucleon wave function, two to describe the anomalous magnetic moments of the up and down quarks,  one to allow for an overall  renormalization of the quark charge at very large $Q^2$, and three to describe the vector dominance structure of the four quark form factors.  Three of these parameters were  fixed, leaving 5 to be varied during the fit.

The present paper extends this model to the description of the $\gamma N\to\Delta$ transition.  As required by the model we use the {\it same\/} CQ form factors and nucleon wave function as fixed in Ref.\ I.   The only freedom remaining is the structure of the  $\Delta$ wave function, and in this paper we study the consequences of the assumption that this wave function is a pure symmetric $S$-wave system of three valence CQ with total spin and isospin equal to 3/2 (described with two parameters).  We conclude that although the electromagnetic form factors of the nucleon alone may be described with such a simple ansatz based on orbital $S$-waves only, the simultaneous description of all the 
$\gamma N \to \Delta$ multipole transition form factors 
demands partial waves with $L>0$. 
This is in agreement  with the findings of
the first non-relativistic quark models
which yield a magnetic dipole (M1) form factor but 
no contributions to 
the electric (E2) and Coulomb (C2) quadrupole form factors 
\cite{Pascalutsa06b}, since they did not
include single quark $D$-states, either in the nucleon or in the $\Delta$ 
wave function. 

In Section II we introduce the $\Delta$ wave function, after a short 
redefinition of the nucleon wave function. In Section III the issues related to
the basis for the polarization states are briefly reviewed. In Section IV the current corresponding to
the N$\to \Delta$ electromagnetic transition is  introduced, and in Section V the form factors are calculated. Section VI presents the results. Section VII
summarizes the work and draws conclusions.

\section{Relativistic $S$-wave functions for baryons}

\begin{figure}[t]
\centerline{\mbox{\includegraphics[width=2.0in]{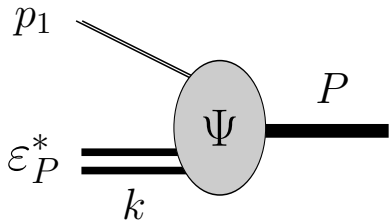}}}
\caption{\footnotesize{
Baryon-quark wave function defined in Eq.\ (\ref{eqNvertex}).  }}
\label{figBaryonVertex}
\end{figure}

 
In the spectator framework 
\cite{nucleon,Gross69,Gross82,Gross92,Stadler97a,comment,Savkli01,Stadler97b,
Gross04b,Adam97,Gross87},
a general baryon with four-momentum $P$ and mass $M$ 
is described by a wave function for an off-shell quark  
and an on-mass-shell diquark-like cluster, defined by 
\be
\Psi(P,k)=(m_q-\not\! p_1)^{-1} 
\bra{k} \Gamma  \ket{P},
\label{eqNvertex}
\ee
where $\Gamma$ is the vertex function describing the coupling of
an incoming on-shell baryon with mass $M$ to an outgoing off-shell
quark and an on-shell quark pair (the {\it diquark}).  
The quark pair is non-interacting with a continuous mass distribution varying from $2m_q$ to infinity.  All of the matrix elements in the spectator theory involve an integral over this mass distribution, and for simplicity this integral is approximated by fixing this mass at a mean value,  $m_s$ (a parameter of the theory that scales out of the form factors, but can be fixed by deep inelastic scattering, as described in Ref.\ I).  The diquark four-momentum, $k=P-p_1$, is constrained by its on-mass-shell condition $k^2=m_s^2$.
The quark has dressed mass $m_q$ and four-momentum $p_1$.  This wave function is shown diagrammatically in Fig.\ \ref{figBaryonVertex}.

Following Refs.\  I and \cite{comment}, we write
the baryon states
in terms of the quark spin  and ``fixed-axis'' diquark 
polarization states, labeled by $\varepsilon_P^\mu$. 
These vectors
were discussed in detail in Ref.~I and \cite{comment}. 
We use them, instead of the diquark helicity vectors, since the latter depend not only on the magnitude
of the diquark momentum, but also on its
direction,  while here we want to consider 
$S$-state orbital effects alone. 
Considering the ``fixed-axis'' polarization states  
we have shown \cite{nucleon,comment}  
that the wave function 
for the nucleon 
transforms as a Dirac spinor
under a Lorentz operation. 
This implies that
the model is covariant
and assures the 
covariance of the electromagnetic current elements.
The method was tested for the nucleon alone in Ref.\ I; here we test it for the $\Delta$.

\subsection{$S$-wave nucleon wave function}

The manifestly covariant $S$-wave nucleon (with mass $m$) wave function was introduced in Ref.\  I
\ba
\Psi_{N\,\lambda_n} (P,k)&=&
\frac{1}{\sqrt{2}} \psi_{N}(P,k) \, \phi^0_I \, u(P,\lambda_n) 
\label{psiRel} \nonumber\\
&- &\frac{1}{\sqrt{6}} \psi_N(P,k) \; 
\phi^1_I \, 
 \gamma_5 \not\!\varepsilon^\ast_P \,u(P,\lambda_n).  \quad 
\ea
Here  $u(P,\lambda_n)$ is a  four-component Dirac spinor, and $\psi_N$
is a scalar function that specifies 
the relative shape of both spin-isospin (0,0)
and spin-isospin (1,1) diquark components.  In the following we will sometimes simplify the notation by suppressing reference to the polarization $\lambda_n$ of the nucleon, and write the nucleon spinor as $u(P,\lambda_n)\to u(P)$.  

In Eq.\ (\ref {psiRel}), 
$\phi_I$ gives the isospin states of the quark-diquark system
($I=\pm 1/2$ is the isospin projection of the nucleon) 
\ba
 \phi_I^0&=&\xi^{0*} \chi^{I} \label{iso0}\nonumber \\
 \phi^1_{I}&=&-\sfrac{1}{\sqrt{3}} \tau \cdot \xi^{1*} \chi^{I}, 
 \label{iso1}
\ea
where $\xi^i$  ({$i=0,1$})  represents the two diquark isospin states,  and  
\ba
\chi^{+\frac12}=\left(\begin{array}{c} 1\cr 0\end{array}\right)=n\,   \quad
\chi^{-\frac12}=\left(\begin{array}{c} 0\cr 1\end{array}\right)=p\ .  \quad
\ea
More details can be found in Ref.\ I. 

The  spin-1 component of the wave function
depends on the fixed-axis
diquark polarization vectors  
$\varepsilon_{\lambda\,P}=
(\varepsilon_0,\varepsilon_x,\varepsilon_y,\varepsilon_z)$,  with $\lambda=0,\pm 1$
the polarization index. 
The explicit expressions for $\varepsilon_{\lambda\,P}$  are given in the next Section 
(Eqs.\  (\ref{eqEPS}) for $m_H=m$). 
Here we emphasize that the polarization vectors 
are written in terms of the nucleon momentum $P$, 
instead of the diquark momentum $k$. 
This dependence gives $\Psi_N$ the correct 
non relativistic limit  
and also assures that $\Psi_N$ satisfies the 
Dirac equation \cite{nucleon}. 

 Note that  (\ref{psiRel}) is written
in terms of  $\varepsilon_P^\ast$, allowing  the interpretation of $\Psi_N$  as an amplitude for an
incoming nucleon and an {\it outgoing\/}  diquark in the final state 
(see Fig.\ \ref{figBaryonVertex}).
 
In this paper we only consider transitions from the nucleon to the $\Delta$ in which the diquark remains a spectator, with its spin and isospin unchanged during the transition.  Since the diquark in the $\Delta$ must have spin-isospin quantum numbers (1,1) [as discussed below], only the spin-isospin (1,1) component of the nucleon wave function is needed.  
It is convenient to introduce a new notation for this component, and rewrite the (1,1) component of Eq.\  (\ref{psiRel}) as
\ba
\hspace{-.5cm}
\Psi_{N\lambda_N}(P,k)&\to&\Psi^{(1,1)}_{N\lambda_N}(P,k)\nonumber\\
&=& - \sfrac{1}{\sqrt{2}} \psi_N (P,k) \phi_I^1 \,  
\varepsilon^{\alpha \ast}_P \, 
U_\alpha(P,\lambda_N),
\label{eqforus1}
\ea
where $\alpha$ is a vector index, and
\be
U_\alpha (P,\lambda_N) \equiv  \sfrac{1}{\sqrt{3}}  \gamma_5 
\left(\gamma_\alpha - \frac{P_\alpha}{m} \right) u(P,\lambda_N).
\label{eqUa}
\ee
Because $\varepsilon_P \cdot P = 0$, this definition is equivalent to the one presented in \cite{nucleon}.
 However, the spinor (\ref{eqUa}) has the properties
\bea
P^\alpha \,U_\alpha (P,\lambda_N)&=&0
\nonumber\\
(m-\not\!P)\,U_\alpha (P,\lambda_N)&=&0
\eea
which are very convenient for the actual calculation of the matrix elements of the
electromagnetic current.   Note also that 
\bea
(m-\not\!P)\,\Psi_{N\lambda_N}(P,k)=0\, .
\eea

The part of the wave function that depends on the magnitude of the relative momentum is modeled by the two parameter function used in Ref.~I:
\be
\psi_N(P,k)=\frac{N_0}{m_s(\beta_1+\chi_N)(\beta_2+\chi_N)}\, .
\label{eqpsiSN} 
\ee
Here $N_0$ is 
a normalization constant 
and $\beta_i$ with $i=1,2$ 
are range parameters in units of $m m_s$.  All of these parameters were fixed in Ref.~I, and we use the same values in this paper.
The dimensionless  variable $\chi_N$ is defined as 
\be
\chi_N=
\frac{(m-m_s)^2-(P-k)^2}{m m_s}.
\label{eqchiN} 
\ee
Since the baryon and the diquark are both on-shell,      the wave function (\ref{eqpsiSN}) can depend only on the  variable $(P - k)^2$ (as required by the Hall-Wightman theorem).
In Appendix \ref{app:G} we show that Eq.~(\ref{eqpsiSN}) assures the asymptotic behavior for the nucleon form factors $G_E$ and $G_M$ will be $1/Q^4$ times logarithmic corrections as expected from perturbative QCD (pQCD). 

This formalism is not restricted to the 
$S$-wave case presented here. It can be 
extended  to states with higher orbital angular 
momentum.  This will be the subject of future work.

The wave function (\ref{eqforus1})  has a very simple physical interpretation.  It is a spin 1/2 state composed of spin 1 and spin 1/2 constituents.  This  spin content is discussed in detail in Appendix \ref{app:A}.

\subsection{$S$-wave $\Delta$ wave function}

The $S$-state wave function for the
$\Delta$ (with mass $M$) is defined similarly to (1,1) 
component of the nucleon wave function
in Eq.\  (\ref{eqforus1}) above.
Non-relativistically a pure $S$-wave  spin 3/2 Delta 
state  can be written as a direct  product
of  spin 1/2 quark and a spin-1 diquark.  
Many details of the nonrelativistic 
construction of the $\Delta$ wave, 
and its relativistic generalization, 
are discussed in Appendix \ref{app:B}.  
In this section we summarize the results.

%

In parallel to Eq.~(\ref{eqforus1}),  the $S$-state Delta covariant wave-function can be written  
\be
\Psi_{\Delta \lambda_\Delta} (P,k)= -
\psi_\Delta (P,k)   \,
\tilde \phi_{I'}  \,
\varepsilon^{\beta \ast}_P  \,
w_\beta(P,\lambda_\Delta), 
\label{eqPsiSD} 
\ee
where $\tilde \phi_{I^\prime}$ is the 
isospin part of the $\Delta$ wave function 
(including a diquark with isospin 1, and playing the same role as the nucleon isospin function $\phi_{I}^1$) with the isospin projections 
$I^\prime=\pm 1/2$ or $\pm 3/2$, $w_\beta(P,\lambda_\Delta)$ 
is the spin 3/2 Rarita-Schwinger 
spinor-vector with spin projections $\lambda_\Delta=\pm1/2,\pm3/2$, and $\psi_\Delta(P,k)$ is a scalar wave function.  
In parallel with the nucleon definition  (\ref{eqforus1}), we define the wave function with a minus sign.
For notational simplicity, the spin indices of the diquark have been omitted from Eq.~(\ref{eqPsiSD}) implying that  $\varepsilon^\alpha_{\lambda P}\to\varepsilon_P^\alpha$ with $\lambda=-1,0,+1$. The  $\Delta$ wave function therefore consists of  three components corresponding to the three different diquark polarizations.  These polarizations are summed in the calculation of the transition form factors, as discussed below.  The diquark spin  polarization vector is, as in the nucleon case,
a function of the $\Delta$ mass and 
momentum.  It will be given explicitly in  Eqs.\ (\ref{eqEPS}) with $m_H =M$.

Since $w_\beta(P,\lambda_\Delta)$ are Rarita-Schwinger  spinor-vectors, they satisfy the usual constraint conditions \cite{Rarita41,Milford55}
\bea
(M-\not \! P)w_\beta(P,\lambda_\Delta)&=&0\nonumber\\
P^\beta w_\beta(P,\lambda_\Delta) &=&0\nonumber\\
\gamma^\beta w_\beta(P,\lambda_\Delta)&=&0\, .
\label{eq:12}
\eea
Therefore, the wave function (\ref{eqPsiSD}) 
satisfies  the Dirac equation 
\bea
(M-\not \! P)\Psi_{\Delta \lambda_\Delta} (P,k)=0\, ,
\eea
showing that  $\Psi_{\Delta \lambda_\Delta} (P,k)$ has no lower (negative energy) components in its rest system.


The isospin wave function $\tilde \phi_{I\prime}$
can be written as 
\be
\tilde \phi_{I^\prime} = \left(
T\cdot \xi^{1 *} \right) \tilde \chi^{I^\prime} 
\label{eqDInr}
\ee
where $\xi^1$ is the  isospin vector of the diquark (identical to the one used for the nucleon),
$\tilde \chi^{I^\prime}$ is a $4\times1$ isospin state, and
$T^i$ denotes the 
$2\times 4$ matrix 
corresponding to the 
$3/2 \rightarrow$ $1/2$ isospin transition operator.  The specific form of these operators is given in Appendix \ref{app:B}.

As in the nucleon case, $\psi_\Delta$ can be expressed as a function of
\be
\chi_\Delta=
\frac{(M-m_s)^2-(P-k)^2}{M m_s}.
\label{eqchiD}
\ee
In particular, we use phenomenological ansatz
\be
\psi_\Delta(P,k)=
\frac{N_1}{m_s(\alpha_1+\chi_\Delta)(\alpha_2+\chi_\Delta)^2},
\label{eqpsiSD1}
\ee
where $\alpha_i$ ($i=1,2$) are 
range parameters in units of $M m_s$ and $N_1$ 
a normalization constant.  We note that the power of $\chi_\Delta$  in the denominator
differs from the corresponding one in the phenomenological ansatz for the nucleon case.
In Appendix \ref{app:G} we show that this choice for the $\Delta$ wave function, 
together with the nucleon wave function (\ref{eqpsiSN}),  give 
the expected $1/Q^4$ pQCD limit for 
the dominant form factors of the electromagnetic
$N \to \Delta$ transition.

\subsection{Covariant spin projection operators}

Both the nucleon and $\Delta$ wave functions have a generic structure
\bea
\Psi_{H\lambda_H}=\phi_{H}(P,k)\;
\varepsilon_P^{\alpha\,*}V_{H\alpha}(P,\lambda_H) \label{eq:17}
\eea
where $H=N, \Delta$, and 
\bea
V_{H\alpha}(P,\lambda_H)&=&\begin{cases} U_\alpha(P,\lambda_N) &\quad {\rm nucleon} \cr
w_\alpha(P,\lambda_\Delta) &\quad\Delta \end{cases}\nonumber\\
\phi_H(P,k)&=&\begin{cases} -\sfrac{1}{\sqrt{2}}\psi_N(P,k)\phi_I^1 & {\rm nucleon}\cr
-\psi_\Delta(P,k) \tilde \phi_{I'} & \Delta\, . \end{cases}
\label{eq:18}
\eea
Furthermore, in both cases
\bea
P^\alpha \,V_\alpha(P,\lambda_H)=0\, .
\label{eq:19}
\eea
These similarities allow us to make some interesting general statements about the nucleon and $\Delta$ wave functions.

The condition (\ref{eq:19}) means that, in the rest frame of the hadron, the vector $V_\alpha$ has spatial components only.  In this subspace the identity operator is 
\bea
{\bf 1}^{\alpha}{}_{\beta}\equiv \widetilde g^{\alpha}{}_{\beta}=g^{\alpha}{}_{\beta}-\frac{P^\alpha P_\beta}{m_H^2}
\eea
where we use the notation $m_N=m$ and $m_\Delta=M$.   
This subspace is spanned by two projection operators:
\be
{\cal P}_{1/2}^{\alpha\beta} + {\cal P}_{3/2}^{\alpha\beta}=\widetilde g^{\alpha\beta}\, ,
\ee
where, using the notation
\bea
\widetilde \gamma^\alpha\equiv \gamma^\alpha-\frac{\not\! P P^\alpha}{m_H^2}
\eea
the two operators are
\bea
&&{\cal P}_{1/2}^{\alpha \beta}={\cal P}_{1/2}^{\alpha \beta}(P)=
\sfrac{1}{3}\; \widetilde \gamma^\alpha \widetilde \gamma^\beta \nonumber\\
&&{\cal P}_{3/2}^{\alpha \beta}={\cal P}_{3/2}^{\alpha \beta}(P)=
\widetilde g^{\alpha \beta}-
\sfrac{1}{3}\; \widetilde \gamma^\alpha \widetilde \gamma^\beta.
\label{eq:23}
\eea
In Appendix \ref{app:C} we show that these operators are  relativistic generalizations of the spin $1/2$ and spin $3/2$ projection operators (for a particle of mass $m_H$).  They operate only in the $3\times3$ subspace of space-like vectors $\widetilde v^\alpha=v^\alpha-P^\alpha (P\cdot v)/m_H^2$.  These operators are well known in the literature, in others contexts (the operator ${\cal P}_{1/2}$ 
is sometimes denoted ${\cal P}_{11}$) \cite{Benmerrouche89,Haberzettl98}.

As expected, the nucleon and $\Delta$ wave functions are eigenvectors of the spin 1/2 and spin 3/2 operators:
\bea
&{\cal P}^{\alpha \beta}_{1/2}\,U_\beta = U^\alpha  \qquad& {\cal P}^{\alpha \beta}_{3/2}\,U_\beta = 0\nonumber\\
&{\cal P}^{\alpha \beta}_{1/2}\,w_\beta = 0 \; \qquad& {\cal P}^{\alpha \beta}_{3/2}\,w_\beta = w^\alpha\,.
\eea
and the orthogonality of these projection operators 
%
\bea
{\cal P}^{\mu \alpha}_{1/2}\,{\bf 1}_{\alpha\beta}{\cal P}_{3/2}^{\beta\nu}= 0 
\label{eq:25}
\eea
implies that the two wave functions are orthogonal, as expected.  We will see later that a generalization of this condition is useful in proving current conservation in the electromagnetic $N \to \Delta$ transition process.

In work already underway \cite{work} these operators will also play an important role in the extension of the formalism to $D$-wave states, for both the nucleon and the $\Delta$.

\section{Fixed-axis diquark polarizations}

\subsection{Definition of the state vectors}

The spin-1 fixed-axis polarization vectors  were introduced in Ref.~I, and some of their properties discussed and derived in \cite{comment}.  (These  are really {\it axial\/} vectors, but for simplicity we will drop the word ``axial'' in the subsequent discussion.)  These  vectors are denoted $\varepsilon_{\lambda P}^\mu$, where $\lambda=0, \pm1$ is the spin projection in the direction of the baryon three-momentum, ${\bf P}$, with $P=\{{\cal E}_p, {\bf P}\}$ the baryon four-momentum and ${\cal E}_p=\sqrt{m_H^2+{\bf P}^2}$ the baryon energy.  In the baryon rest frame the fixed axis may be chosen to be in any direction.  Choosing the  $\hat z$ direction the polarization vectors are 
\ba
\varepsilon_{0 P_0}^\mu &=& 
\left( 0, 0, 0, 1 \right) \nonumber \\ 
\varepsilon_{\pm P_0}^\mu &=& \sfrac{1}{\sqrt{2}} 
\left( 0, \mp 1, - i,0 \right). 
\label{eqEPS0}
\ea
where the baryon four-momentum in its rest frame is  denoted $P_0=\{m_H,0,0,0\}$.  These polarization vectors, when used in Eqs.~(\ref{eqforus1}) and (\ref{eqPsiSD}),
give the correct non-relativistic limit 
for the nucleon and $\Delta$ wave function. 

To write  the baryon wave function  in a frame
where  the baryon is moving, a boost  
in the direction of motion  ($z$-direction by convention)  of the baryon is needed.
For this choice, $P=({\cal E}_{\rm P},0,0,\mbox{P})$, and the polarization vectors become
\ba
\varepsilon_{0 P}^\mu &=& \frac{1}{m_H}
\left({\mbox{P}}, 0, 0, 
{\cal E}_{\rm P}\right) \nonumber \\ 
\varepsilon_{\pm P}^\mu &=& \sfrac{1}{\sqrt{2}} 
\left( 0, \mp, -i,0 \right)\, ,
\label{eqEPS}
\ea
and satisfy
\be
\varepsilon_{\lambda P}^\ast \cdot \varepsilon_{\lambda^\prime P}= -
\delta_{\lambda \lambda^\prime}, \hspace{0.8cm}
\varepsilon_{\lambda P}\cdot P =0.
\label{eqEPSdot}
\ee

In the following we will
use the variable $P_-$ for the momentum in the initial state 
and $P_+$ for the momentum in the final state.
Also, the initial state will be  a nucleon 
and the final state will refer either to 
nucleon or to a $\Delta$.

In the Breit frame,  for a transition from
an initial state of mass $m$ to a final 
state of mass $M$, with a 
momentum transfer $q=P_+-P_-$,  we write 
\ba
& &
P_-=\left(E_-,0,0, -\sfrac{1}{2} q_L\right)
\nonumber\\
& &
P_+=\left(E_+,0,0, \frac{1}{2} q_L\right), 
\label{eqPp}
\ea
where $E_+=\sqrt{M^2 + \sfrac14 q_L^2}$ 
and $E_-=\sqrt{m^2 + \sfrac14 q_L^2}$ with 
\be
q_L^2 = Q^2 
+ \frac{(M^2-m^2)^2}{2(M^2+m^2)+Q^2}.
\ee
Eqs.\ (\ref{eqPp}) hold for both equal masses 
$M=m$ ($q_L=\sqrt{Q^2}$) 
and unequal masses. 

In the Breit frame, according to Eqs.\  (\ref{eqEPS}), 
we have for the initial state
\ba
& &\varepsilon_{0 P_-}^\mu = \frac{1}{m}
\left(-\sfrac{1}{2}q_L, 0, 0, 
E_- \right) \nonumber \\ 
& &
\varepsilon_{\pm P_-}^\mu = \frac{1}{\sqrt{2}} 
\left( 0, \mp 1, -i,0 \right), 
\label{eqEPSpm}
\ea
and for the final state
\ba
& & \varepsilon_{0 P_+}^\mu = \frac{1}{M}
\left(\sfrac{1}{2} q_L, 0, 0, 
E_+ \right) \nonumber \\ 
& &\varepsilon_{\pm P_+}^\mu = \frac{1}{\sqrt{2}} 
\left( 0, \mp 1, -i,0 \right). 
\label{eqEPSpp}
\ea
Note that the polarization vectors
$\varepsilon_{P_\pm}^\mu$ from Eqs.\ 
(\ref{eqEPSpm})-(\ref{eqEPSpp}) refer to
the Breit frame only.


Starting from the Breit frame
we can then change to an arbitrary frame 
by  means of a suitable
Lorentz transformation $\Lambda$.  The details of this transformation are discussed in Ref.~\cite{comment}.  The Breit frame momentum $P^\mu$ (with $P=P_\pm$)  
is then transformed  into $P^{\prime \mu}$ 
according to 
\be
P^{\prime \mu}= \Lambda^\mu_{\; \nu} P^\nu.
\label{eqEPP}
\ee
Due to the four-vector character of $\varepsilon_P^\mu$, 
the polarization vectors in the new 
frame $\varepsilon_{P^\prime}^\mu$, 
parametrized by $\Lambda$, 
become
\be
\varepsilon_{\lambda P^\prime}^\mu =
\Lambda^\mu_{\; \nu} \varepsilon_{\lambda P}^\nu,
\label{eqEPSp}
\ee
for each polarization $\lambda$. 
In this notation 
only the momentum index distinguishes the arbitrary 
frame (momentum $P^\prime$) 
from the Breit frame (momentum $P$) 
polarization vector.  

The {\it same} transformation $\Lambda$
acts on both final and initial momenta $P_\pm$.
The transformation (\ref{eqEPSp}) 
combined with the transformation law of 
the Dirac spinors and 
the Rarita-Schwinger states (see Appendix B)  
implies that the baryon wave functions 
transform as Dirac spinors.
The demonstration of the same property 
for the $\Delta$ wave function follows the 
lines of the presented in the Ref.~\cite{comment}.

Fixed-axis  polarization vectors $\varepsilon_P$  are different from the helicity vectors used  in Ref.\
\cite{GrossAga},  which we denote here by $\eta$. 
The latter depend on the diquark momentum $k$, and therefore
on its direction, satisfying
$\eta\cdot k=0$. 
The helicity vectors $\eta$ can be related to our fixed-axis polarization states by a rotation \cite{comment}.
In fact, with 
an appropriate redefinition of the vertex function $\Gamma$,  a wave function using fixed-axis states can be made exactly equivalent to another wave function using helicity states. 
In the case of an initially totally spherical symmetric
wave function, the transformation from diquark fixed-axis polarization states to direction-dependent or helicity states
gives a vertex function accompanying the helicity vectors just the right
angular dependence on the diquark momentum to cancel the dependence introduced  by  the helicity states $\eta$ \cite{comment}. 
 Conversely, a spherically symmetric vertex function $\Gamma$, like the one used here, if taken together with the direction-dependent diquark
 helicity states $\eta$, would result in a wave function without  spherical symmetry. Since here we want to investigate  the consequences
 of spherical symmetric wave functions only, it becomes natural to write these wave functions in terms of fixed-axis diquark polarizations.

\subsection{Importance of the collinearity condition}

 
We emphasize that matrix elements of states that include fixed-axis polarization vectors must {\it first\/} be constructed in a frame in which the incoming and outgoing states have {\it collinear\/} three-momenta, and only after this has been done can the matrix elements be transformed to an arbitrary frame.  If matrix elements are constructed in this order, they will be both unique and covariant, but if they are not constructed in this order, they will be neither covariant nor unique.  A simple example of the problems encountered if one does not start with a collinear frame is developed in Appendix \ref{app:D}.  Our failure to emphasize this point in our original presentation of these ideas lead to a criticism of Kvinikhidze and Miller \cite{Miller07a}, which we addressed completely in Ref.~\cite{comment}.

Why is a collinear frame required? We will see in the next section that the matrix elements we calculate assume that the diquark is a spectator which does not participate in the interaction.  Hence,  that the polarization of the diquark emitted by the initial baryon must be the same as the polarization of the diquark absorbed by the final baryon.  There are not two distinct diquarks, but one 
and only one, diquark.  Therefore, using fixed-axis polarization states, we must be certain  that the polarization states of the diquark 
emitted from the initial vertex,
and of the diquark absorbed into the final vertex, are defined with
respect to the {\it same axis\/}, and only in the collinear frame are we certain that the definitions
of the polarization of the incoming and outgoing diquark (the {\it same\/} diquark)
are consistent with a single direction.  Therefore, if we happen to be presented with an interaction in which the initial baryon three-momentum ${\bf P}_-$ is not parallel to the final baryon momentum, ${\bf P}_+$, we must first transform the matrix element of the collinear frame, construct the matrix element, and then transform back to the original frame.

Fortunately, given a any initial and final
momentum configuration, there is always a
Lorentz transformation that will boost
and rotate both states to a collinear frame with the three-momenta ${\bf P}_\pm$ in the $z$ direction, so the need to define the states in collinear frame imposes no limitation. 
Our construction is similar to the definition of two-particle helicity states 
in the two-body center-of-mass by Jacob and Wick \cite{JacobWick}.  

\section{matrix elements of the current}

\subsection{Definition of the current}

\begin{figure}[t]
\centerline{
\mbox{
\includegraphics[width=3.0in]{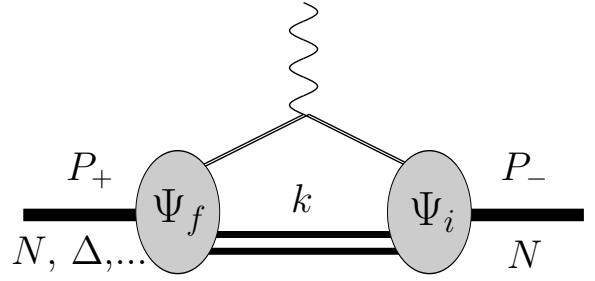}}}
\caption{\footnotesize{
Relativistic impulse approximation.}}
\label{impulse}
\end{figure}

Consider the electromagnetic transition from an initial state $\Psi_i$  (mass $m$) and a 
final state $\Psi_f$ 
(mass $M$).  The relativistic impulse approximation (RIA) to the transition current  in the spectator formalism, shown in Fig.~\ref{impulse},  is
\be
J_{fi}^\mu (P_+,P_-)=
3 \sum_\lambda 
\int_k \overline \Psi_f (P_+,k) 
j_I^\mu \Psi_i(P_-,k).
\label{eqCur}
\ee
This is covariant, but we will study it in the collinear Breit frame where the fixed-axis polarizations are consistently defined, as discussed above, and the incoming and outgoing momentum, $P_\mp$, are already defined in Eq.~(\ref{eqPp}).

In the RIA, the photon couples to {\it each\/} quark through the diagram shown in Fig.~\ref{impulse}.  The factor of 3 in (\ref{eqCur}) comes from  isospin invariance, which allows us to express the sum of the three diagrams in terms of a single integral  multiplied by 3.  All intermediate states are taken into account by summing over the diquark spin-1 polarizations and integrating over all 
positive on-mass-shell diquark (spectator) states 
with energy $E_s$, using 
\be
\int_k= \int 
\frac{d^3 k}{(2\pi)^3 2 E_s}.
\label{intk}
\ee
There is, in principal, an integration over all spectator masses, $m_s$, but this integral is replaced by the value of the integrand at some (unknown) mean value $m_s$, which becomes a parameter of the model.

The quark current $j_I^\mu$ (which is isospin dependent) is decomposed into its
Dirac and Pauli terms, 
\be
j_I^\mu= j_1 \gamma^\mu +
j_2 \frac{i \sigma^{\mu \nu} q_\nu}{2m},
\label{eqQuarkC}
\ee 
where $j_i$  ($i=1,2$) are the quark form factors, 
defined in Ref.~I.

These form factors $j_1$ and $j_2$ 
include the quark structure and
can be decomposed into isoscalar and isovector components:
\ba
& &j_1= \frac{1}{6} f_{1+} (Q^2) + \frac{1}{2}f_{1-} (Q^2) \tau_3 \nonumber\\
& &j_2= \frac{1}{6} f_{2+} (Q^2) + \frac{1}{2}f_{2-} (Q^2) \tau_3\, ,
\ea
where $\tau_3$ is the quark isospin operator.
In this work we adopted the quark form factors from Ref.\ \cite{nucleon}
\ba
f_{1\pm} (Q^2) &= &
\lambda + \frac{(1-\lambda)}{1+Q^2/m_v^2} +
\frac{c_{\pm} Q^2/M_h^2}{\left(1 + Q^2/M_h^2 \right)^2} \qquad
\nonumber\\ 
f_{2\pm} (Q^2) &= &
k_{\pm} 
\left\{ 
\frac{d_\pm}{1+Q^2/m_v^2} 
+
\frac{(1-d_{\pm})}{1+Q^2/M_h^2} 
\right\}\, . \label{eqf1m}
\ea
In these expressions $m_v$ and $M_h$ are  
vector meson masses that represent the dominant contributions from the vector 
dominance model (VDM).
The lower mass, $m_v= m_\rho$ (or $m_\omega$), 
describes of the two pion resonance 
(three pion resonance) effect 
and $M_h$, fixed as $2m$ (twice the nucleon mass), 
takes account of all the large mass resonances.

The parameter $\lambda$ is  
fixed by the deep inelastic scattering (DIS)
distribution amplitudes and can interpreted physically 
as a scaling of the quark charges in DIS limit.
All the other parameters are presented in Table \ref{Nucleon_table}.
They correspond to a previous application
of the covariant Spectator
theory to  the description of the nucleon form factor data
with only an $S$-state in the nucleon wave function 
 \cite{nucleon}. 
We know from the beginning that the restriction to orbital $S$-waves  is a 
considerable simplification, 
baryon ground state to its first resonance, 
but it is interesting to see exactly 
what are the consequences of such a simple assumption.

Note that all of these parameters are fixed 
by the nucleon data: elastic form factors   
and DIS.
The  quark electric form factors $f_{1 \pm}$ are  
normalized to 1 at $Q^2=0$, in order to reproduce the quark and
nucleon charge.
The quark magnetic form factors are normalized  
by proton and neutron magnetic moments 
$f_{2+}(0)=\kappa_+= 1.639$ and $f_{2-}(0)=\kappa_-= 1.823$; 
see Ref.\  \cite{nucleon} for a detailed discussion.

\begin{table}[t]
\begin{center}
\begin{tabular}{c c c c c  c}
Model   & $\beta_1$, $\beta_2$ & $c_+, c_-$ & 
$d_+,d_-$ & $\lambda, m_s/m$ &  $N_0^2, \chi^2$ \\
\hline 
 I     &  0.057      &  2.06 &  -0.444       
       &  1.22       &  10.87 \\
       &  0.654      & 2.06 &  -0.444    
       &  0.88       & {\bf 9.26} \\
\hline
II     &  0.049      &  4.16  & -0.686
       &  0.547      &  11.27\\   
       &  0.717      &  1.16  & -0.686
       &  0.87       & {\bf 1.36}  \\ 
\hline
\end{tabular}
\end{center}
\caption{Parameters of the nucleon wave function 
($\beta_1,\beta_2$ and $N_0^2$) and quark form factors.
In each case we kept $\kappa_+= 1.639$ and 
$\kappa_-=1.823$ in order to reproduce  
the nucleon magnetic moments exactly.
The difference in the $\chi^2$ is mainly due 
to the description of the neutron electric
form factor. Model I preserves the 
isospin symmetry for the quark electric form factor 
$f_{1+}=f_{1-}$, but cannot describe 
neutron electric form factor data.
See details in Ref.\  \cite{nucleon}.} 
\label{Nucleon_table}
\end{table}

Due to the relation $T_i^\dagger \tau_i=0$ only the 
isovector components of the current contributes 
to the $\gamma N \to \Delta$ transition.
Then, in the following discussion we need the isovector current only,
which can be written
\bea
j_I^\mu\Big|_{\rm v}= j_v^\mu\, \frac{\tau_3}{2} =\Big\{f_{1-} \gamma^\mu +
f_{2-} \frac{i \sigma^{\mu \nu} q_\nu}{2m}\Big\}\, \frac{\tau_3}{2},
\label{eqQuarkC2}  
\eea
with the isovector quark form factors $f_{1-}$ and  $f_{2-}$ 
defined as above.


\subsection{Diquark polarization sum}

The next step in the general reduction of the transition current is to carry out the sum over the diquark polarizations.  Using the generic notation of Eq.~(\ref{eq:17}), the current is written 
\begin{align}
J_{fi}^\mu (P_+,P_-)
=& \frac{3}{2}  \int_k\Big[ \phi_f(P_+,k)\,\tau_3\,\phi_i(P_-,k)\Big]
D^{\beta \alpha}
\nonumber\\
 &\times\overline V_{f\beta} (P_+,\lambda_+)\, j_v^\mu \,V_{i\alpha}(P_-,\lambda_-)\, ,
 \label{eq:43}
\end{align}
where we assume that only the isovector quark current contributes to form factor (true for the $\gamma N\to\Delta$ transition) and have written the diquark spin sum as the operator
\be
D^{\mu \nu} \equiv
\sum_\lambda \varepsilon_{\lambda P_+}^\mu 
\varepsilon_{\lambda P_-}^{\nu \ast},
\label{eqDBF}
\ee
that is evaluated in Ref.~\cite{comment} and Appendix \ref{app:E}.  The final result shows that $D^{\mu \nu}$ depends only on the momenta and masses of the two states, and can be written 
\ba
D^{\mu \nu}
& =&-\left(g^{\mu \nu} 
- \frac{P_-^\mu P_+^\nu}{P_+ \cdot P_-} \right) 
 \label{eqDMm}\\
&+&a \left(P_-^\mu -\frac{P_+ \cdot P_-}{M^2} P_+^\mu \right)
    \left(P_+^\nu -\frac{P_+ \cdot P_-}{m^2} P_-^\nu \right),
\nonumber 
\ea
where the factor $a$ is
\be
a=-\frac{Mm}{P_+ \cdot P_-
\left[ M m + P_+ \cdot P_- \right]}\, .
\label{eqAdef}
\ee
Note that  $D^{\mu\nu}$ satisfies the conditions
\be
P_{+ \mu} D^{\mu \nu} =  D^{\mu \nu} P_{- \nu} =0\, .
\ee

\subsection{Current conservation}

Current conservation requires that $q_\mu J_{fi}^\mu =0$.  To see if this condition is satisfied, we consider separately
the Dirac current (from the quark charges) and the Pauli current 
(from the quark anomalous magnetic moments).
The Pauli current is always conserved,
independent of the asymptotic states considered.
To reduce the  Dirac current we use the facts that 
the initial and final states both satisfy the 
Dirac equation, and that the charge form factors 
of the quark depend on  $q^2$ and can be factored out  of the integral 
\ba
q_\mu J_{fi}^\mu&=& \sfrac{3}{2} f_{1-} 
\sum_\lambda 
\int_k \overline \Psi_f \,\tau_3
\not \!q\; \Psi_i \nonumber \\
&=& \sfrac{3}{2}(M-m) f_{1-}\int_k \overline \Psi_f \, \tau_3 \Psi_i
\label{eqCC}
\ea 
If the masses are equal, the condition is automatically satisfied, but for unequal, masses the states must be orthogonal
%
\be
\sum_\lambda \int_k \overline \Psi_f \,\tau_3 \Psi_i=0.
\label{eqOrtog}
\ee
We can also write Eq.\  (\ref{eqCC}) 
using the notation of Eq.~(\ref{eq:43}) 
\begin{align}
q_\mu J_{fi}^\mu&=
\frac{3}{2}(M-m) f_{1-} \int_k\Big[ \phi_f(P_+,k) \tau_3\,\phi_i(P_-,k)\Big]
\nonumber\\
 &\qquad\times\overline V_{f\beta} (P_+,\lambda_+)\,D^{\beta\alpha} 
\,V_{i\alpha}(P_-,\lambda_-)\,.
\end{align}

For the $\gamma N\to\Delta$ transition, we can use the projection operators to prove orthogonality. Using the fact that the $N$ and $\Delta$ states are eigenvectors of the spin-1/2 and spin-3/2 projection operators, we can write
\begin{align}
&\overline V_{\Delta\beta} (P_+,\lambda_+)\,D^{\beta\alpha} \,V_{N \alpha}(P_-,\lambda_-)\\
&=\overline V_{\Delta \mu} (P_+,\lambda_+)\Big[ {\cal P}^{\mu\beta}_{3/2}(P_+)\,D_{\beta\alpha}{\cal P}^{\alpha\nu}_{1/2}(P_-) \Big]\,V_{N \nu}(P_-,\lambda_-)\, , \nonumber
\end{align}
where ${\cal P}_{3/2}(P_+)$ and ${\cal P}_{1/2}(P_-)$ are the projection operators of Eq.~(\ref{eq:23}) with $P\to P_+$ and $P\to P_-$, respectively.
We show in Appendix \ref{app:C} that the operator in square brackets is zero:
\bea
{\cal P}^{\mu\beta}_{3/2}(P_+)\,D_{\beta\alpha}{\cal P}^{\alpha\nu}_{1/2}(P_-)=0\, . \label{eq:51}
\eea
This is the generalization of the orthogonality relation (\ref{eq:25})
and proves the orthogonality of the wave functions for all momentum transfers, $q$.  
Due to this orthogonality between the initial and the final states,
the additional \mbox{-$\not\!q q^\mu/{q^2}$} term used in the definition of the current in Ref.~I 
vanishes in this application. This is why we did
not include that extra term in Eq. (\ref{eqQuarkC}).

\section{$\gamma N \to \Delta$ transition} \label{sec:V}

We will now apply the formalism of the previous sections to  
the study of the electromagnetic $N \Delta$ transition 
which has a simple 
interpretation in terms of valence quark structure:
the $\Delta$ is a result of a spin flip of a single quark
in the nucleon.
It is then understandable that the
magnetic dipole multipole M1 dominates 
the transition for low $Q^2$, while
the electric E2 and the Coulomb C2 quadrupoles 
give contributions of only a few percent.
For large $Q^2$ however, according to perturbative QCD  
estimations, we expect 
equal contributions from M1 and E2 \cite{Carlson}.
At the present the $Q^2$ scale of the pQCD 
regime is not known, which motivates calculations within models.

\subsection{Simplification of the transition current}
\label{VC}

The transition current (\ref{eq:43}) can now be simplified.   
Using the notation of Eqs.~(\ref{eq:18}) 
and (\ref{eqDMm}) and changing the integration variable 
from $k$ to $k/m_s$ we obtain an integral
independent of the diquark mass $m_s$ (due 
to the wave functions normalization which goes with $1/m_s$) 
and the diquark energy factor $E_s/m_s=\sqrt{1+k^2/m_s^2}$
(which altogether cancel the 
$m_s^3$ dependence of the element $d^3 k$).  
Factoring out the isospin factors gives 
\begin{align}
J_{\Delta N}^\mu (P_+,P_-)
=&-\sfrac3{2\sqrt{2}}(\tilde\phi_{I'})^\dagger \tau_3\phi_I^1 \int_k\Big[ \psi_\Delta(P_+,k)\psi_N(P_-,k)\Big]\;
\nonumber\\
 &\times\overline w_{\beta} (P_+,\lambda_+)\, j_v^\mu \,U_{\alpha}(P_-,\lambda_-)D^{\beta\alpha}\, ,
\label{eqJiso}
\end{align}
where $j_v^\mu$ is the defined in terms of the 
isovector part of the quark current (\ref{eqQuarkC2}):
\be
 j_v^\mu = f_{1-} \gamma^\mu +
f_{2-}  \frac{ i \sigma^{\mu \nu} q_\nu}{2m}\, ,
\label{eqJtil1}
\ee
which is the only part of the quark current to contribute to the transition amplitude.  The isospin matrix element is evaluated using the properties of the isospin matrix transition $T^i$ 
(between spin 1/2 states and 3/2 states).   
Summing over the isospin projections $m_{\footnotesize I}$ of the diquark isospin vector gives
\bea
(\tilde \phi_{I^\prime}^1)^\dagger \tau_3 \phi_I^1&=&  -\sfrac{1}{\sqrt{3}}\tilde \chi^{I'\dagger} T^\dagger_i \tau_3
    \tau_j  \chi^I  \sum_{m_{\footnotesize I}}\xi^1_i(m_{\footnotesize I})\xi^{1*}_j(m_{\footnotesize I})
\qquad\nonumber\\
    &=& -\frac{1}{\sqrt{3}}\tilde \chi^{I'\dagger} \Big(T^\dagger_i \tau_3
    \tau_i \Big) \chi^I  =-\frac{2 \sqrt{2}}{3}  \delta_{I I^\prime}\, .
\eea
Next, using the fact that the initial and final states both satisfy the Dirac equation, we may reduce the Pauli form of the current using the Gordon decomposition
\bea
 \frac{ i \sigma^{\mu \nu} q_\nu}{2m}=\gamma^\mu \Big(\frac{M+m}{2m}\Big)-\frac{P^\mu}{m}
\eea
where $P^\mu$ is the average of the initial and final momentum, defined in Eq.~(\ref{eq:54}).  We already saw in the discussion of gauge invariance [leading up to the identity (\ref{eq:51})] that the matrix element of the identity operator is zero, and hence the $P^\mu$ term does not contribute.  This allows us to collect the quark charge and anomalous magnetic moment contributions into a single term, giving finally
\begin{align}
J_{\Delta N}^\mu (P_+,P_-)
=&\;\delta_{I'I}\,f_v\int_k\Big[ \psi_\Delta(P_+,k)\psi_N(P_-,k)\Big]\;
\nonumber\\
 &\times\overline w_{\beta} (P_+,\lambda_+)\, \gamma^\mu \,U_{\alpha}(P_-,\lambda_-)D^{\beta\alpha}\, ,
\label{eq:69}
\end{align}
where 
\be
f_v=f_{1-}+\frac{M+m}{2m}f_{2-}\, ,
\label{eqJM2}
\ee
is a particular linear combination of the quark from factors that can be factored out of the integral because it depends on $Q^2$ only.

Equation (\ref{eq:69}) includes the explicit 
conservation of the  $z$-projection of the isospin.
This means that the model predicts
that the amplitude is the same for both isospin channels: 
$\gamma^\ast p \to \Delta^+$ and $\gamma^\ast n \to \Delta^0$.


The nucleon wave function $\psi_N$ 
is normalized to one,  as required 
by the charge conservation at $Q^2=0$ \cite{nucleon}.
Similarly, also the $\Delta$ wave function 
(\ref{eqPsiSD}) 
is constrained, in the rest frame 
where $\bar P=(M,0,0,0)$, by  
the charge condition 
(excluding the isospin states from the wave function):
\ba
Q_I&=&
3 \sum_\lambda \int_k 
\overline \Psi_\Delta (\bar P,k) j_1 \Psi_\Delta (\bar P,k)
\nonumber \\
&=& 
 \left(
\frac{1+\overline T_3}{2}
\right) 
\int_k |\psi_\Delta (\bar P,k)|^2,  
\label{eqQdel}
\ea
where $j_1=\sfrac{1}{6} + \sfrac{1}{2} \tau_3$. 
The isospin operator $ \overline T_3$ 
is defined as 
\be
\overline T_3=
3 \sum_{i} T_i^\dagger \tau_3 T_i=
\left[
\begin{array}{rrrrrr} 
       &  3 & & 0 & 0 & 0 \cr 
       &  0 & & 1 & 0 & 0 \cr
       &  0 & & 0 &-1 & 0 \cr
       &  0 & & 0 & 0 &-3 \cr 
\end{array}
\right].
\ee
Equation (\ref{eqQdel}) 
gives the correct $\Delta$ charge if 
\be
\int_k |\psi_\Delta (\bar P,k)|^2=1. 
\ee
This condition determines also  the normalization constant 
for the wave function in Eq.\  (\ref{eqpsiSD1}).

\subsection{Form factors: Generalities}

The $N \to \Delta$ electromagnetic  transition current 
(excluding the electron charge $e$ and ignoring the polarizations of the nucleon and the $\Delta$) is given by 
\be
J^\mu = \overline w_\beta(P_+) \Gamma^{\beta \mu} 
(P,q)\gamma_5 u(P_-),
\ee
where the general form of the transition vertex $\Gamma^{\beta \mu}$ is 
\be
\Gamma^{\beta \mu } (P,q) =  q^\beta \gamma^\mu G_1+ q^\beta P^\mu G_2 + 
 q^\beta q^\mu G_3-  g^{\beta \mu}G_4\, .
\label{eqJS}
\ee 
The variables $P$ and $q$ are respectively the average of 
baryon momenta and the photon momentum: 
\ba
P&=&\sfrac12(P_++P_-)\nonumber \\
q&=&P_+-P_-. \label{eq:54}
\ea
The form factors $G_i$, $i=1,..,4$ depend only on 
$Q^2=-q^2$.
Due to current conservation, $q_\mu \Gamma^{\beta \mu} =0$, only 
three of the four form factors are independent. 
In particular, we can write $G_4$ in terms of the first three form factors  
\be
G_4=(M+m) G_1 +\sfrac12(M^2-m^2) G_2 -Q^2 G_3,
\label{eqG4}
\ee
and adopt the structure originally proposed by 
Jones and Scadron \cite{Jones73}.

The parametrization (\ref{eqJS}) is not 
directly comparable to experimental data.
More convenient for that purpose are the magnetic dipole (M), 
electric quadrupole (E) and Coulomb quadrupole (C) 
form factors defined as 
\begin{align}
G_M^\ast(Q^2)= \kappa\,\Big\{ 
&\left[(3M +m)(M+m) +Q^2 \right] \frac{G_1}{M} \nonumber \\
& +(M^2-m^2) G_2 -2 Q^2 G_3 \Big\} \\
G_E^\ast(Q^2)=
\kappa\,\Big\{&(M^2 -m^2 -Q^2) \frac{G_1}{M} 
\nonumber\\
& +  
(M^2-m^2) G_2 -2 Q^2 G_3 \Big\}
 \\
G_C^\ast(Q^2)=\kappa\,
\Big\{&4M G_1 +(3M^2+m^2+Q^2)G_2   
\nonumber\\ 
&+ 
2 (M^2-m^2-Q^2) G_3   
\Big\}\, ,
\end{align}
where
\be
\kappa=\frac{m}{3(M+m)} \, .
\ee
The three form factors $G_a^\ast$ ($a=M,E,C$) 
are related to the magnetic, electric, and Coulomb (or scalar)
multipole transitions, respectively.

\subsection{Form Factors: Application of the model}

Substituting for $U_\beta$ in (\ref{eq:69}), 
and suppressing the isospin conservation factor $\delta_{I^\prime I}$, 
gives immediately
\begin{align}
J_{\Delta N}^\mu (P_+,P_-)
=&\overline w_{\beta} (P_+,\lambda_+)\, {\cal O}^{\beta\mu}\gamma^5\, u(P_-,\lambda_-)\, ,
\label{eq:71}
\end{align}
where
\bea
{\cal O}^{\beta\mu}=\sfrac{1}{\sqrt{3}}f_v \int_k\Big[ \psi_\Delta(P_+,k)\psi_N(P_-,k)\Big] \gamma^\mu D^{\beta\alpha}\gamma_\alpha\, .\label{eq:72}
\eea

This is easily reduced; the work is given in Appendix \ref{app:F}.  The final results for the form factors 
of a transition between $S$-wave nucleon 
and  $\Delta$ states are
\ba
& &G_M^\ast (Q^2)= 
\frac{8}{3 \sqrt{3}} \frac{m}{(M+m)}\, f_v\; {\cal I}  
\label{eqGMSS}\\
& &G_E^\ast (Q^2)= G_C^\ast (Q^2)=0,
\label{eqGESS2}
\ea 
where 
\be
{\cal I}=
\int_k 
\psi_\Delta(P_+,k) \psi_N(P_-,k)\, .
\ee
Note that ${\cal I}$ is the only factor which depends on the 
scalar wave functions and it  is Lorentz scalar 
(frame independent).

\section{Results}

Before we present the numerical results we focus on
the analytical structure of the results (\ref{eqGMSS})-(\ref{eqGESS2}).
Note that 
the electric and Coulomb quadrupole form factors vanish in this calculation, 
which is restricted to $S$-wave orbital states.
This result is consistent with quark models 
based on quark $S$-wave states \cite{Pascalutsa06b}. 
According to the literature, 
the presence of multipoles E2 and C2 
is a signature of the nucleon and/or $\Delta$ deformation.
Our result of identically vanishing  quadrupole form factors
is the consequence of considering only  $S$-states
for both nucleon and $\Delta$ wave functions, and consequently
a nucleon and a $\Delta$ with spherical form. 
The inclusion of higher orbital momentum components 
would generate non-vanishing  
electric and Coulomb quadrupoles, as we will confirm in forthcoming work.
It is worth noticing that the experimental results for the 
transition multipoles indicate a contribution 
of E2 and C2 of the order of a few percent 
at $Q^2=0$, consistent with a small angular momentum component in the wave function.

Next, look at the magnetic form factor, $G_M^\ast$,
at $Q^2=0$.   Substituting for $f_v$ gives
\be
G_M^\ast (0)=
\frac{2}{3 \sqrt{3}} 
\left(\frac{2 m}{M+m} + \kappa_- \right) {\cal I}(0).
\label{eqGM0a}
\ee
The isovector magnetic moment $\kappa_-$ was fixed 
by the nucleon magnetic moment in Ref.\ \cite{nucleon}
\be
\kappa_-=
\sfrac{3}{5} (\mu_p - \mu_n) -1=1.823,
\label{eqKm}
\ee 
For future discussion we write Eq.\ (\ref{eqGM0a}) as 
%
\be
G_M^\ast (0)= \frac{2}{\sqrt{3}} 
\left[ 
\frac{\mu_p -\mu_n}{5} 
- \frac{1}{3} \frac{M-m}{M+m}
\right] {\cal I}(0).
\label{eqGM0b}
\ee
When the experimental nucleon magnetic 
moment is used in (\ref{eqGM0a}) 
one has
\be
G_M^\ast (0)=
2.07 \, {\cal I}(0).
\label{eqGMmax}
\ee
where 
\be
 {\cal I}(0)= \left. \int_k \psi_\Delta \psi_N \right|_{Q^2=0}.
\ee
Note, however, that we are working with 
normalized wave functions 
\be
\left. \int_k |\psi_N |^2\right|_{Q^2=0}=1, 
\hspace{1cm} 
\left. \int_k |\psi_\Delta |^2\right|_{Q^2=0}=1. 
\label{eqNorma}
\ee 
Because of this conditions, the integral ${\cal I}(0)$ 
is limited, in its absolute value, by the H\"{o}lder inequality,
the version of the 
Cauchy-Schwartz inequality for an integral
\be
\left|\int_k \psi_\Delta \psi_N \right| \le 
\sqrt{\int |\psi_\Delta|^2} \sqrt{\int |\psi_N|^2}. 
\ee
In particular for $Q^2=0$ we obtain 
\be
|{\cal I}(0)| \le 1.
\ee
Choosing the positive sign (consistent with our 
definitions of the scalar wave functions) 
we conclude that
\be
G_M^\ast (0) \le 2.07.
\label{eqUnGM}
\ee
Since the experimental value is considerably larger than this limit,
$$ G_M^\ast (0)=  3.02 \pm 0.03,$$
we see that the  Spectator quark model,  
in impulse approximation, can, at best, only describe 69\% 
of the $\gamma N \to \Delta$ transition form factor $G_M^\ast$ 
at the photon point.

This underestimation of $G_M^\ast(0)$ is an universal property of all constituent quark models.
It has been previously reported in the literature 
\cite{Donoghue75,Isgur82,Warns90,Capstick94,Bijker94,JDiaz04,JDiaz05}. 
Naive $S$-wave non-relativistic quark 
models with  $SU(6)$ symmetry (and no dynamical effects included)  \cite{Pascalutsa06b}, predict
\be
G_M^\ast (0)= \frac{2 \sqrt{2}}{3} 
\sqrt{\frac{m}{M}} \mu_p = 2.30 \, .
\label{eqGMnqm}
\ee
Including kinematic effects, Ref.\ \cite{JDiaz07a} obtains
\be
G_M^\ast (0)= \frac{2 \sqrt{2}}{3} 
\frac{\sqrt{2 E_N(m+E_N)}}{m+M} \sqrt{\frac{m}{M}}
\mu_p = 2.04,
\label{eqGM0c}
\ee 
where $E_N$ is the nucleon energy at threshold
in the $\Delta$ rest frame.
For a review of the constituent quark models 
predictions see Ref.\ \cite{Pascalutsa06b}.

Our model
for a quark-diquark system differs from other quark models; 
our quarks are not static, as 
in Eq.\ (\ref{eqGMnqm}), and our  magnetic form factor 
is related to both $\mu_p$ and $\mu_n$,
not only to $\mu_p$ as in Eq.\  (\ref{eqGM0c}).
Nevertheless, comparing Eqs.\  (\ref{eqGM0b}) and 
(\ref{eqGM0c}) we can see that very different descriptions 
can lead to the similar results if the same constraints 
are considered (normalization of the wave functions). 
For completeness we add that calculations based on QCD sum rules also lead to an 
underestimation of $G_M^\ast(0)$ although
these models do not apply to the $Q^2=0$ region 
\cite{Braun06a}. 
Similar results are obtained using 
Generalized Parton Distributions (GDP)
 \cite{Stoler,Guidal05,Pascalutsa06d}.
These models extrapolate the Parton Distribution 
Functions from Deep Inelastic Scattering 
to intermediate energies.
For low $Q^2$ $G_M^\ast$ is underestimated 
by 20-30\%  \cite{Guidal05,Pascalutsa06d}.

The failure of quark models 
to describe the $\gamma N \to \Delta$ transition at threshold 
shows their limitations, which stem from taking constituent 
quarks  as the only 
relevant degrees of freedom.
Quark wave functions can be normalized to correctly describe 
nucleon and $\Delta$ static charges, but fail in the description 
of the dynamical $\gamma N \to \Delta$ transition which 
does not involve a charge density, but instead 
a transition charge density.
The magnitude of the difference 
between the quark model result, 
which we label  the {\it Bare} result,
and the experimental result, may be
 due to pion field contributions, 
and  is a manifestation of the 
strong correlation between the $\Delta$ 
and the $\pi$N system.

\subsection{Decomposition into Bare and Pion Cloud form factors}

\begin{figure}
\vspace{1.0cm}
\centerline{
\mbox{
\includegraphics[width=8cm]{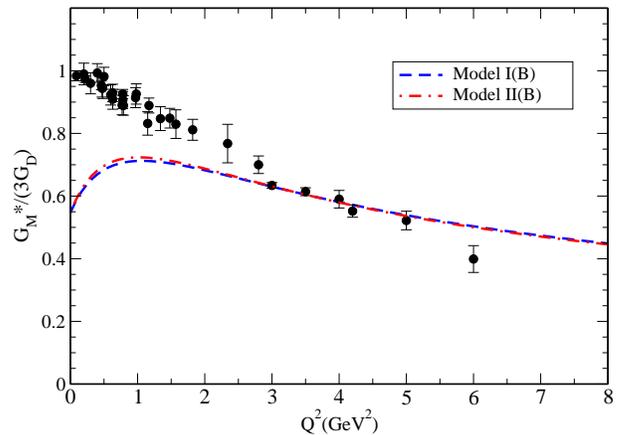} }}
\caption{Result of the fit the $\Delta$ 
wave function parameters to the data $Q^2 \ge 2.9$ GeV$^2$ 
(where $G_M^\pi$ is expected to be very small). 
The nucleon and quark parameters are given 
by model I and II (see Table \ref{Nucleon_table}).
Delta parameters are presented in Table \ref{Delta_table}.
Data from CLAS/Jlab \cite{CLAS02,CLAS06}, 
DESY \cite{Bartel68} and SLAC \cite{Stein75}. }
\label{figBareFit}
\end{figure} 

Following the previous discussion, 
we decompose  $G_M^\ast$ into  two contributions:
\be
G_M^\ast (Q^2)=G_M^B(Q^2)+G_M^\pi(Q^2).
\label{eqGMexp}
\ee
The term $G_M^B$ is the Bare form factor: 
the contribution of the quark core given 
by the quark model under consideration.
The term $G_M^\pi$ is a contribution 
due to the pion field:  the contribution from any diagram involving a
photon and  pion loops.
Our  spectator quark model can predict   $G_M^B$ only.

There are two kinds of descriptions that 
take into account the effects 
of the pion field explicitly: dynamical models  and low momentum 
Effective Field Theories or  
Chiral Perturbation Theories.  
Here we focus on dynamical models
\cite{SatoLee,Kamalov,Pascalutsa}
because these models can be used to describe the entire
momentum region over which  data is available.

A dynamical model uses  hadronic 
degrees of freedom and a coupled channel method to derive  transition amplitudes 
involving initial and final  
meson-baryon and photon-baryon states.
The transition amplitude can be decomposed 
in two components: (i) the background 
or non-resonant amplitude which is 
the solution of an Lippmann-Schwinger-like  
equation with a non-resonant interaction kernel;
and (ii) the resonant amplitude which includes the contributions from dressed  intermediate baryon resonance states.
The non-resonant interaction kernel that generates the 
non-resonant background includes direct couplings 
of the photon or mesons with the baryons, and may also 
include meson rescattering described by intermediate vector mesons. 
The resonant part is generated by dressing the $s$-channel pole terms 
generated by vertex functions describing the couplings of photons 
or mesons to the bare baryon pole.
Both the non-resonant direct coupling terms and the resonance vertex functions  
are parametrized by simple phenomenological expressions 
with parameters adjusted to fit the pion-nucleon 
and pion photo-production data.
%
%
A review can be found in Refs.\  
\cite{Pascalutsa06b,Burkert04a,Drechsel06a}.

The effective contribution 
of the pion cloud depends on the 
particular model.  The models of Sato and Lee (SL) \cite{SatoLee},  
Dubna-Mainz-Taipei (DMT) \cite{Kamalov} 
predict that the pion cloud will give an important 
contribution at $Q^2=0$  that falls quickly with increasing $Q^2$.
The Utrecht-Ohio model \cite{Pascalutsa} in opposition 
predicts small contributions for $Q^2 \approx 0$ 
and more significant contributions of $Q^2 \sim 2 $ GeV$^2$ 
for the pion cloud.
According to Ref.\ \cite{Burkert05}, pion cloud 
effects give 33\% of the total contribution at  $Q^2=0$, 
and less than 10\% for $Q^2 > 4$ GeV$^2$.

Pure quark models include no pion cloud effects 
and can give a complete description only at higher  
$Q^2$ where contributions from the  pion cloud are negligible.
Supporting this interpretation, 
our numerical calculations show that we can fit 
the region $Q^2> 2.5$ GeV$^2$, but not the low $Q^2$ region.
A fit to the higher $Q^2$ data ($Q^2 \ge 2.9$ GeV$^2$) is presented 
in Fig.\ \ref{figBareFit}. 
The $\Delta$ wave function parameters obtained from the two fits shown, referred to as Models I(B) and II(B), are given in Table  \ref{Delta_table}.
We omitted from  this Table  the parameters of the nucleon wave function model which also enters the calculation, 
since
that wave function was already fixed by the nucleon form factors and DIS results  \cite{nucleon},  and those parameters
were shown already in Table \ref{Nucleon_table}.


From the figure we conclude that 
with no explicit pion cloud 
we can explain about 55\% of $G_M^\ast$ at $Q^2=0$.
This contribution is lower 
that the upper limit of Eq.\ (\ref{eqGMmax}).
For each model, the theoretical quantity ${\cal I}(0)$  is a measure of the extent to which a model approaches its theoretical
upper limit, and the deviation of $G_M^\ast(0)/3$ from the experimental value  1 is a measure of the  quality of the fit to the data at $Q^2=0$.

\begin{table}
\begin{center}
\begin{tabular}{c c c c c}
Model   & $\alpha_1$, $\alpha_2$ & $\lambda_\pi$, $\Lambda_\pi^2$ & 
   $N_1,{\cal I}(0)$ & $G_M^\ast(0)/3, \chi^2$ \\
\hline 
 I(B) &  0.169      & $ -$ &   2.88       
       &  0.548 \\
       &  0.489      & $ -$ &  0.792    
       &  {\bf 1.32}\\
\hline
II(B) &  0.181      & $-$  &  3.05
       &  0.547 \\   
       &  0.493      & $-$  &  0.790
       &  {\bf 1.26} \\ 
\hline 
I & 0.313 & 0.474 & 2.88  & 1.026\\
  & 0.374 & 1.172 & 0.798 & {\bf 2.64} \\
\hline
II& 0.290 & 0.464 & 2.95  & 1.012  \\
  & 0.393 & 1.224 & 0.794 & {\bf 1.84} \\
\hline
\end{tabular}
\end{center}
\caption{The dimensionless Delta wave function parameters $\alpha_1$ and $\alpha_2$ and the normalization constant  $N_1$ are defined in  Eq.\   (\ref{eqpsiSD1}).
Model I(B) and II(B) include no pion cloud; Models I and II include a pion cloud with parameters 
$\lambda_\pi$ and $\Lambda_\pi^2$ (in GeV$^2$) defined in Eq.\ (\ref{eqGpi}).  The overlap integral between nucleon and Delta wave-function 
(which cannot be larger than 1) is
${\cal I}(0)$; 
$G_M^\ast(0)/3$ measures the quality of the fit 
for $Q^2=0$ (where the experimental result $\simeq 1$). }
\label{Delta_table}
\end{table}

\begin{figure}
\vspace{1.0cm}
\centerline{
\mbox{
\includegraphics[width=8cm]{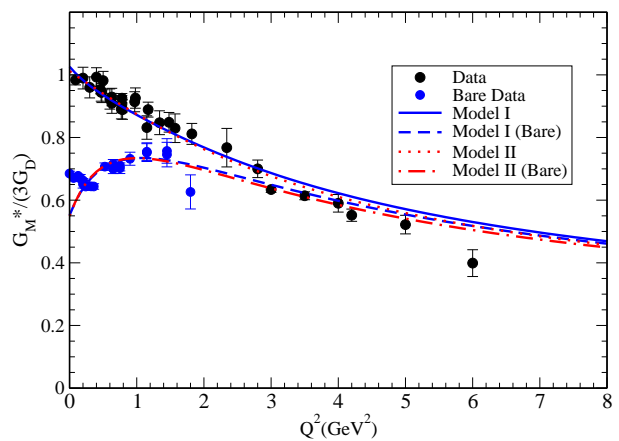}}}
\caption{Fit to the $G_M^\ast$ and 
Bare data using using model I and II.
The Bare data is from \cite{JDiaz07a,JDiaz07b}, 
$G_M^\ast$ data from figure \ref{figBareFit}.
The Bare result is now defined considering $\lambda_\pi=0$ 
in the dressed models.}
\label{figBarePion}
\end{figure}

To compare our model with the data over the entire range of $Q^2$, 
we need a parametrization for $G_M^\pi$. 
Based on the magnitude of the effects in the DMT 
model and particularly in the SL model  \cite{Burkert05},
we used a very simple double dipole approximation for 
the pion cloud
\be
\frac{G_M^\pi}{3 G_D} = \lambda_\pi 
\left(\frac{\Lambda_\pi^2}{\Lambda_\pi^2+Q^2} 
\right)^2,
\label{eqGpi}
\ee
where $G_D=1/(1+Q^2/0.71)^2$ (with $Q^2$ in GeV$^2$)  and  
$\lambda_\pi$ and $\Lambda_\pi^2$ are parameters
to be adjusted to the data.   The parameter  $\lambda_\pi$
can be interpreted 
as the fraction of pion cloud effects at $Q^2=0$ and $\Lambda_\pi^2$ measures 
the falloff of the pion cloud.
Note that we  parametrize the ratio of $G_M^\pi$ to
$3G_D$ following the tradition of 
scaling $G_M^\ast$ with the dipole factor $G_D$.
This form assume a  falloff of  $G_M^\pi \sim 1/Q^8$, to be 
compared with $G_M^\ast \simeq 3 G_D \sim 1/Q^4$.
%
%

It is important to realize that, due to the 
complexity of the pion production 
process, the decomposition of Eq.\  (\ref{eqGMexp}) 
is strongly model dependent. 
Dynamical models can differ in coupling constants, 
off-mass-shell extrapolations, and off course 
the parametrization of  bare component itself.
Using Eq.\ (\ref{eqGpi})  we can find 
several combinations $(G_M^B,G_M^\pi)$ with 
approximately the same sum $G_M^\ast$.  
We must find some way to constrain one of these two components.

To do this, we use a procedure implemented for the first time in Ref.\ \cite{JDiaz07a}.  
Using the SL model, Julia-Diaz, Lee, Sato and Smith extract,
independently at each $Q^2$ point, a value of the bare form factor.  This is possible because the bare form factor, $G_M^B$, is one of the parameters that enters the dynamical SL model, and it is therefore possible to determine it, without any theoretical bias, by a best fit to the data.

The data for the bare component of the form factor, determined in this way, are shown in Fig.~\ref{figBarePion}.   Note that a separation between the "bare" data and the experimental data can only be made in the region $Q^2 < 2.5$ GeV$^2$.
Above $Q^2\sim$ 3 GeV$^2$ the pion cloud contributions are 
less significant and the dynamical model produces only 
a correction to the bare component; 
in this region the full contribution comes 
mainly from the bare component. 

Models I and II, dressed by the pion cloud, are a result 
of a simultaneous fit of both components of the form factor, 
Eq.\ (\ref{eqGMexp}), to the $G_M^\ast$ experimental data 
\cite{CLAS02,CLAS06,Bartel68,Stein75} 
and of the bare component,  
to the bare ``data'' for $Q^2<2$ GeV$^2$ based 
on the SL extraction \cite{JDiaz07a,JDiaz07b} discussed above.  
The parameters for Models I and II are compared to models I(B) 
and II(B)  in Table \ref{Delta_table}.
(The $\chi^2$ given there is for the fit to $G_M^\ast$ only.)  
In Fig.~\ref{figBarePion} we present the results our 
predictions for the dressed models and the respective 
Bare version obtained setting $\lambda_\pi=0$.
In the same figure we can see that
the data below 0.13 GeV$^2$ (first three points) 
cannot be described by our model.
This limitation is related to the behavior of 
the overlap integral of the nucleon and Delta wave functions.  Note that the values of ${\cal I}(0)$ are almost identical for the bare and dressed models, but the dressed models now have  $G_M^\ast(0)/3 \approx 1$ 
due to the addition of a pion cloud term 
of about 46\% at $Q^2=0$.
Furthermore, a reasonable 
description of the 'bare data' 
is obtained for both models ($\chi^2=4.2$ for model I and $\chi^2=4.6$ for model II) at least for $Q^2>0.13$ GeV$^2$.
 
It is worth to mentioning that the parametrization 
of the dressed models changes the results of the bare contribution
relative to the Fig.~\ref{figBareFit}.
Although similar, 
the bare results presented in Fig.~\ref{figBarePion} 
are slightly larger than the results of   
the models I(B) and II(B)  presented in Fig.~\ref{figBareFit},
in particular for $Q^2 < 2$ GeV$^2$.
This increment is not obvious in 
the graphs, but as we can see in table \ref{Delta_table}, 
the parameters $\alpha_1$ and $\alpha_2$ for the 
models I(B) and II(B) 
are significantly different from the models I and II. 
This feature is the result of including the 
low momentum ``bare data'' ($Q^2< 2$ GeV$^2$) that appears 
not be completely consistent with the 
high $Q^2$ data ($Q^2> 3$ GeV$^2$) at least 
with the {\it naive} parametrization (\ref{eqGpi}). 
To sort out this situation more high $Q^2$ data of high quality are needed (the current data set includes  only 6 data points with $Q^2 \ge 3$ GeV$^2$, compared with 26 data points for   $G_M^\ast$ and 21 data points 
for $G_M^B$ for $Q^2<3$ GeV$^2$).

\subsection{Comparing with other models for the Bare form factors}

As mentioned above, the dynamical models need 
a phenomenological parametrization of  
the ``bare'' vertex.
%
There is some freedom in the choice of the ``bare'' form factor, but the constraints of the quark models 
are usually taken into account.
This parametrization can be done for each
transition multipole $G_\alpha^\ast$ ($\alpha=M,E,C$).
For the SL and DMT models the ``bare'' form factors 
can be written 
\be
G_M^B=G_M^B(0) (1+ a Q^2) \exp(-bQ^2) f,
\label{eqGMBdm}
\ee 
where $G_M^B(0) \le 2.07$ fixes the contribution 
of the quark core for $Q^2=0$, $a, b$ are positive 
parameters, and $f=1$ for the SL model and 
$f = \sqrt{1 + \frac{Q^2}{(M+m)^2}}$ for the DMT model.
For SL $G_M^B(0) = 2$; for DMT $G_M^B(0) =1.65$.
All other parameters can be found 
in Refs.\  \cite{SatoLee,Kamalov}.

The structure of the Utrecht-Ohio model \cite{Pascalutsa}
for the bare form factor is incompatible 
with a pion cloud which is not zero
for $Q^2=0$, and for this reason a direct 
and simple estimation of the bare form factor 
based in Eqs.\  (\ref{eqGMexp}) and (\ref{eqGpi}) 
is not possible.

\begin{figure}
\vspace{1.0cm}
\centerline{
\mbox{
\includegraphics[width=8cm]{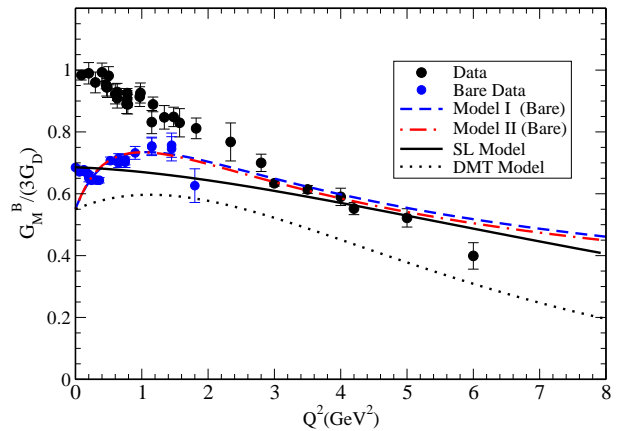} }}
\caption{Comparing Bare form factor parametrization 
with the 'bare data' from Refs.\ \cite{JDiaz07a,JDiaz07b}  
with models I and II, 
SL model \cite{SatoLee} and DMT model \cite{Kamalov}.}
\label{figFullBare}
\end{figure}

The bare form factors use by of SL and DMT [obtained from the analytical expression (\ref{eqGMBdm})]  and our numerical result 
for models I and II are compared in Fig.\
\ref{figFullBare}.  We see that 
the bare SL form factor overestimates 
the data for $Q^2 > 4$ GeV$^2$ and that 
DMT always underestimates the data 
(suggesting a significant pion cloud contribution 
even for $Q^2 > 3$ GeV$^2$, since for 4 GeV$^2$ the effect 
is still about 10\%).
We need to point out that the particular 
parametrization of both SL and DMT,
in particular the SL model,  
was done before the data $Q^2 >  4$ GeV$^2$ became 
available \cite{CLAS06}.
The new data showed the limitation of the 
particular parametrization, and was one of the motivations  
for work presented in Ref.\ \cite{JDiaz07a}, where the Bare form factor are adjusted 
for each $Q^2$ point.  Note that our results are very similar to  SL for $Q^2 > 3$ GeV$^2$.

\subsection{Comparing Model I with Model II}

In this work we consider two different models 
for the nucleon wave function 
as presented in Ref.\  \cite{nucleon}.
We conclude that the model II 
is the best model for the $\gamma N \to \Delta$ transition 
because is the one that better describes 
the high $Q^2$ data 
(meaning the pion cloud contribution is better
described by the dipole form).
Model I gives a  slightly worse description 
of the data, but both models are almost 
indiscernible in the region $Q^2 < 3$ GeV$^2$. 
This result is very interesting
because the models are fundamentally different 
in the description of the nucleon form factors. 
In the model I the isospin symmetry
is exactly imposed with $f_{1+}= f_{1-}$ 
(Eq.\  (\ref{eqf1m}) with $c_+=c_-$),
leading to the failure 
of the description of the neutron 
electrical form factor (see Ref.\  \cite{nucleon}) 
and a consequent high $\chi^2$ penalization 
as shown in table 
\ref{Nucleon_table}.
As the $\gamma N \to \Delta$ transition is independent 
of $f_{1+}$ (only isovector form factors contribute)
a reasonable description of the form factors 
can be obtained for both models.

\subsection{Discussion}

Since our model  includes only $S$-waves in the nucleon and 
$\Delta$ wave functions, the 
comparison of our results to all models and 
frameworks used in the description 
of the electromagnetic $N \to \Delta$ transition has to be done with care.
As mentioned above, we can only predict non-zero contributions to the dominant 
form factor $G_M^\ast$ (55\% of it for $Q^2=0$),
while the experiments reveal two more  
non-vanishing, albeit small, form factors ($G_E^\ast$ and $G_C^\ast$).
The limitation of our model is visible in its failure  
at high $Q^2$, where $G_E^\ast$
is comparable with $-G_M^\ast$, according to pQCD.
That regime is however out of reach 
of the present state-of-the-art measurements.

On the other hand,
our results are hardly comparable with 
the low momentum Effective Field Theory 
and Chiral Perturbation Theory 
\cite{Pascalutsa05,Pascalutsa06c,Faessler06,Ramirez07,Gail06}. 
Those models include pion degrees of 
freedom consistently at least 
at one pion loop level. 
But
the range of the prediction is limited to 
$Q^2<$ 0.25 GeV$^2$,  due to the expansion in terms of the terms 
of the small variables (pion momentum, pion mass,
difference of $\Delta$ and nucleon mass) .
Besides the range limitation, the bare 
contribution is adjusted to the data around
$Q^2=0$ and is not really predicted from 
quark structure. This 
leads to a bare contribution significantly
different from quark models.
Note that  Effective Chiral Perturbation Theory  
relies on an energy scale parameter, usually $\lambda \sim 1$ GeV,
decoupling the short range physics (bare) 
from the long range physics where the pion cloud is included.
In Ref.\  \cite{Faessler06} the pion cloud gives
a positive contribution; in Ref.\  \cite{Gail06} 
the contribution is negative (bare contribution 
$G_M^B(0)=4.04$).

As for lattice QCD data, the comparison  
is not yet conclusive:
the (quenched) lattice data of Alexandrou {\it et al} \cite{Alexandrou04} 
overpredict $G_M^\ast$ in the chiral limit
for $Q^2 >$ 0.1 GeV$^2$.
In principle, the quenched lattice data should be comparable to 
the bare form factor results, and hence it might be expected 
to be larger that the experimental data,  
but instead it underestimates the data.  
It is not known yet either  this discrepancy 
is due to the limitations of the quenched data 
or to the extrapolation to the physical region.
The available results from full QCD (unquenched) 
are not adequate for an extrapolation  \cite{Alexandrou07b}.

The soundest comparison to be made, then, is  
to other valence quark models.  This is why in the previous subsection we compared 
the magnitude of the Bare form factor obtained by us
to the results from  dynamical models and constituent 
quark models based on $S$-wave wave functions.
We are left then with comparing our results to the predictions of 
Light-Cone Sum Rules of Braun {\it et al} \cite{Braun06a}.
This formalism divides the main contributions of 
the form factors into  two components: the soft  contribution
falling with $1/Q^6$ and the hard contribution 
due to pQCD with a $1/Q^4$ falloff.
The soft contributions, dominant  
in the intermediate region,   
are explicitly evaluated,
using the nucleon asymptotic amplitudes 
(valence quark distributions),
and an adjustable momentum range parameter (Borel parameter). 
Their results describe $G_M^\ast$ in the  
$3-6$ GeV$^2$ region, but fail in the region of low and high $Q^2$ 
(where they have an almost constant slope).
Also they underestimate $G_M^\ast$ 
at low $Q^2$ like the constituent quark models 
(at $Q^2=1$ GeV$^2$ the prediction for $G_M^\ast$
only takes account of 60\% of the experimental value).

The success of our model in the description 
of the bare form factors (high $Q^2$ region) is related to our
 choice for the form of  $\psi_\Delta$ in Eq.\  (\ref{eqpsiSD1}).
In particular, the extra power in the $\alpha_2+ \chi_\Delta$  
factor, comparatively to the nucleon wave function, plays a key role.
It is also interesting to note that
our results would be very similar to the 
results of the Ref.\   \cite{Braun06a} 
had we used a $\Delta$ scalar function 
of the  nucleon-type  (see Eq.\  (\ref{eqpsiSN})).
This generates $G_M^\ast$ with an almost constant slope and 
$G_M^\ast/(3G_D) \simeq 0.6$.
As mentioned, our parametrization 
for the $\Delta$ scalar form factor $\psi_\Delta$ 
is consistent with  $G_M^\ast \sim 1/Q^4$ for large $Q^2$,
i.e., the pQCD prediction. 
For the $\Delta \Delta$ form factors 
(a $\Delta$ in the initial and final state) 
our  prediction, based on  Eq.\  (\ref{eqpsiSD1}), 
has a  $1/Q^6$ falloff
for the dominant form factors at high  $Q^2$, 
instead  of the  $1/Q^4$ advocated by  pQCD.
We can find a good compromise  for both the
low $Q^2$ description and the expected pQCD behavior, by 
considering 
\be
\psi_\Delta = \frac{a}{(\alpha_1^\prime+\chi_\Delta)(
\alpha_2^\prime+\chi_\Delta)} 
+\frac{b}{(\alpha_1+\chi_\Delta)(\alpha_2+\chi_\Delta)^2},
\ee
where $\alpha_i^\prime$ are new range parameters, 
and $a,b$ coefficients which balance the two regimes: 
$a$ the deep $Q^2$ asymptotic region 
and $b$ the low-intermediate $Q^2$ region.
This compromise requires $b >> a$ 
pushing the pQCD dominance to very far away.

\section{Summary and Conclusions}

We have developed a systematic formalism  for the description of 
 baryon  wave functions, built upon 
constituent quark 
(flavor, spin and isospin) and baryon (spin, isospin) effective properties.
The form of the nucleon and $\Delta$  wave functions
with $S$-wave orbital angular momentum  components only 
is presented.
The formalism is manifestly covariant. 
The wave functions are covariant and transform like  
Dirac and Rarita-Schwinger spinors. 
The matrix elements are covariant, with the same form in all frames.
 
One of our models (model II)  describes the nucleon form factor data 
\cite{nucleon}   and the dominant contribution to  the 
$N \to \Delta$ electromagnetic transition.
The results for  $G_M^\ast$  
show a reasonable agreement with the data, 
and explain its measured falloff.
They are also consistent with 
the long range behavior predicted by  pQCD 
($G_M^\ast \simeq  1/Q^4)$ \cite{Carlson}.
Our results for  $G_M^\ast$ 
are consistent with the results of
quark models where only $S$-states are considered.

In agreement  with previous works (see Refs.\ 
\cite{Pascalutsa06b,JDiaz07a,Burkert04a,Braun06a}),  
we conclude that a successful description of 
$G_M^\ast $ near $Q^2=0$ requires an addition 
of a pion cloud term not included in the class 
of valence quarks 
to explain the strength at $Q^2=0$ of the  magnetic form factor $G_M^\ast$ 
of the $N \to \Delta$ electromagnetic transition.
Our predictions for the pion cloud  
underestimate the predictions based on the 
Sato and Lee model (46\% versus 33\%) for  $Q^2=0$. 
This gap can in principle decrease 
once $D$-states are included into 
the $\Delta$ wave function.
The magnitude of the pion cloud for $Q^2=0$ 
is similar to the estimations of the DMT model 
but we predict a faster falloff.
[Note that our results are not directly compared 
with the DMT model because our 'bare' contribution 
is fixed by the 'bare data' extraction of the SL model.]
Except for the region $Q^2 \sim 0$,
our model is consistent with the 'Bare data'  
extraction based in Sato and Lee model \cite{JDiaz07a}.
For the region $Q^2> 3$ GeV$^2$ our model 
and the original SL model \cite{SatoLee} 
overestimate the data slightly.
In this region either the pion cloud parametrization 
is not adequate or the data is insufficient 
to constraint adequately the  parameters 
of the pion cloud. More higher $Q^2$ data and more accurate data for 
both $G_M^\ast$ and $G_M^B$ (SL model) 
is therefore necessary for 
to establish the effect of the pion cloud.

Next, we plan to generalize
the structure of the wave functions
to include higher orbital angular momentum 
states in the quark-diquark system, without loss 
of the covariance requirement. 
The inclusion of D states in the nucleon and 
$\Delta$ is  in progress.



\begin{acknowledgments}

The authors wants to thank 
to B.\ Juli\'a-D\'{\i}az  for sharing 
the 'bare data' of Ref.\ \cite{JDiaz07a}.
This work was partially support by Jefferson Science Associates, 
LLC under U.S. DOE Contract No. DE-AC05-06OR23177.
G.~R.\ was supported by the portuguese Funda\c{c}\~ao para 
a Ci\^encia e Tecnologia (FCT) under the grant  
SFRH/BPD/26886/2006. 

\end{acknowledgments}


\appendix



\section{Nucleon wave function}
\label{app:A}

In the non-relativistic limit, the wave function of a spin 1/2 
system composed of a quark (spin 1/2) and a quark 
pair (diquark), as in  the nucleon,  can be decomposed into two
components: the scalar part (spin-0 diquark) 
and the axial-vector part (spin-1 diquark).

The spin-0 part is just 
\be
\phi_s^0= \chi_s,
\ee
where $\chi_s$ is the usual Pauli spinor 
($2 \times 1$ state).
The spin of the system is then given by  
the spin of the quark.
The relativistic generalization is $u(P_0,s)$, where $P_0=\{m,{\bf 0}\}$ is the four-momentum of a nucleon at rest.

The spin-1 component of the wave function $\phi_s^1$ 
describes the quark-diquark spin 1/2 system (nucleon) 
in the initial state and a diquark polarization 
vector in the final state.
The quark-diquark spin state, represented by  $V_{Ns}$,
is a direct product 
of a spin-1 
diquark state with a spin 1/2 quark state.
In this case the three-component vector $\varepsilon^i_\lambda$ (with $i=x,y,z$ and $\lambda=0,\pm$  the diquark polarization index) describes the diquark polarization [these vectors are the three-component parts of the polarization vectors defined in  Eqs.\ (\ref{eqEPS0})] .
Then
\be
(V_{Ns})^i =
\sum_{\lambda s^\prime} 
\braket{1\, \lambda; \sfrac{1}{2} s^\prime }{ \sfrac{1}{2} s}
\varepsilon_{\lambda }^i \chi_{s^\prime}, \label{A2}
\ee
where $s=\pm 1/2$ is the spin projection of the nucleon,
$\braket{s_1\, m_1 ; s_1\,  m_2 }{j \, m_j}$
is the Clebsch-Gordan coefficient that couples spins $s_1$ and $s_2$ to total spin $j$, and
$\chi_{s^\prime}$ the quark spinor.
Explicitly 
\ba
 & &
\left(V_{N,+\!\frac{1}{2}}\right)^i =
 \sqrt{\frac{2}{3}} \varepsilon_+^i\, \chi_{-\frac{1}{2}}
-  \sqrt{\frac{1}{3}}  \varepsilon_0^i\, \chi_{+\frac{1}{2}} \nonumber \\
& &
\Big(V_{N,-\!\frac{1}{2}}\Big)^i =
 \sqrt{\frac{1}{3}} \varepsilon_0^i \,\chi_{-\frac{1}{2}}
-  \sqrt{\frac{2}{3}}  \varepsilon_-^i\, \chi_{+\frac{1}{2}}.
\nonumber
\ea
Equation (\ref{A2}) can be written 
\be
(V_{Ns})^i= -\frac{1}{\sqrt{3}} \sigma_i \chi_s,
\ee
where $\chi_s$ now represents the nucleon spinor.
The natural relativistic generalization is
\be
(V_{Ns})^i \to U^\alpha (P_0,s) =  \frac{1}{\sqrt{3}}
\gamma_5 \left(\gamma^\alpha-\frac{P_0^\alpha}{m}\right) u(P_0,s),
\label{eqUs}
\ee 
where $u(P_0,s)$ is the Dirac spinor of 
the nucleon and 
$\alpha=\{0,i\}$ with $U^0=0$.

Note that Eq.\ (\ref{eqUs}) describes only 
the initial state of the vertex
represented in Fig.\  \ref{figBaryonVertex} 
(the 3-quark bound state). 
To obtain the amplitude of the full process we 
need to contract  $(V_s)^i $ with 
the diquark final state $\varepsilon_\lambda^{\alpha \ast}$
(the quark is off-shell with its final spin state unspecified).
As result, we have the amplitude
\be
\phi^1_s = - \varepsilon_{\lambda P_0}^{\alpha \ast} 
U_\alpha(P_0,s) \, .  \label{A5} 
\ee
Equations (\ref{eqUs}) and (\ref{A5})  can both be generalized 
for a moving nucleon by means of a boost in the $z$-direction.



\section{Spin structure for the $\Delta$ $S$-state}
\label{app:B}


In close analogy with the nucleon, the $\Delta$ wave function, in its rest frame rest frame, can be written as a direct product of a spin-1 diquark and a spin-1/2 quark
\be 
(V_{\Delta s})^i =
\sum_{\lambda s^\prime} 
\braket{1\, \lambda; \sfrac{1}{2} s^\prime }{ \sfrac{3}{2} s}
\varepsilon_{\lambda }^i \,\chi_{s^\prime}.
\label{eqPhi32}
\ee
where $s=\pm 3/2$ or $\pm 1/2$ is the spin projection of the $\Delta$.
Once again
$\varepsilon_\lambda^i$ is  the diquark polarization 
in the $\Delta$ rest frame,  
and $\chi_{s^\prime}$ a quark Pauli spinor.

We can also express $V_{\Delta s}$ in terms of a basis of spin $3/2$ states:
\ba
& &
\omega_{+{\frac{3}{2}}}=
\left(\begin{matrix}1\cr0\cr0\cr0\cr\end{matrix}\right)\qquad
\omega_{+{\frac{1}{2}}}=
\left(\begin{matrix}0\cr1\cr0\cr0\cr\end{matrix}\right)\qquad 
\nonumber \\
& &
\omega_{-{\frac{1}{2}}}=
\left(\begin{matrix}0\cr0\cr1\cr0\cr\end{matrix}\right)\qquad
\omega_{-{\frac{3}{2}}}=
\left(\begin{matrix}0\cr0\cr0\cr1\cr\end{matrix}\right).
\ea
In this case the connection between the $\omega$ and $V_{\Delta s}$ can be written
\be
(V_{\Delta s})^i= {\cal T}^i \omega_s,
\ee
where ${\cal T}^i$ is an $2 \times 4$ matrix  
that transforms the spin 3/2 state of the $\Delta$ into the spin 1/2 state (of a quark).
The elements of ${\cal T}^i$ can be  evaluated 
using the coefficients in Eq.\ (\ref{eqPhi32}).
The result is
\ba
{\cal T}^x&=&-\frac{1}{\sqrt{6}}\left(\begin{matrix} \sqrt{3}&0&-1&0\cr
0& 1 &0&-\sqrt{3}\end{matrix}
\right) \qquad \nonumber \\
& & \nonumber \\
{\cal T}^y&=&-\frac i{\sqrt{6}}\left(\begin{matrix} \sqrt{3}&0&\;1\;&0\cr
0&\;1\;&0&\sqrt{3}\end{matrix}
\right) \qquad \\
& & \nonumber \\
{\cal T}^z&=&\sqrt{\frac{2}{3}}\left(\begin{matrix}0& \;1\;&
0&0\cr
0&0&\;1\;&0\end{matrix}
\right)\, .
\nonumber 
\ea
As in Eq.\ (\ref{eqPhi32}), $(V_{\Delta s})^i$  
is a $2 \times 1$ spinor with a spin 1/2 structure.

To convert (\ref{eqPhi32}) to relativistic form, add a negative energy $4 \times 1$ lower component 
that vanishes in the $\Delta$ rest frame:
\be
(V_{\Delta s})^i \to  \left[
\begin{matrix}
{\cal T}^i \omega \cr 0 \cr 
\end{matrix} \right]
\equiv 
w^i(P_0,s).
\label{eqRSext}
\ee 
Here  $w^i(P_0,s)$ is the Rarita-Schwinger vector-spinor
for  3-momentum ${\bf P}=0$.
It satisfies the constraint Eqs.\ (\ref{eq:12}).  These constraints insure that $w^0$ vanishes in the rest frame, as implied by Eq.\  
(\ref{eqRSext}).

To generalize the states to an arbitrary frame with  ${\bf P} \ne 0$,
we boost them using a Lorentz transformation $\Lambda$, giving 
\be
w^{\beta} (P,s) =
S (\Lambda) \Lambda^\beta{}_{\alpha} w^\alpha(P_0,s)\, .
\ee


Using the state $w^{\beta}$ in a arbitrary frame, 
the  $\Delta$-quark-diquark  vertex is constructed 
in the same way as the nucleon vertex.  
Considering  the final state diquark polarization vector, 
$\varepsilon^\ast_\lambda$, following the nucleon state convention of Eq.\ (\ref{A5})
\be
\tilde \phi^1_s = - \varepsilon_{\lambda P_0}^{\beta \ast} 
w_\beta(P_0,s) \,,
\ee
gives the spin wave function introduced of Eq.\ (\ref{eqPsiSD}).



\section{Relativistic Spin Projection operators}
\label{app:C}

In this paper we work with operators ${\cal O}^{\alpha\beta}$ that satisfy the constraint equations
\bea
P_\alpha {\cal O}^{\alpha\beta} =0= 
{\cal O}^{\alpha\beta} P_\beta \label{eq:C1}
\eea
In the particle rest system, such operators ``live'' in the 3$\times3$ subspace corresponding to  nonrelativistic 3 dimensional space, and it is easy to relate these operators to their nonrelativistic analogues.

As an example, consider the projection operators that operate on the direct product of spin-1 and spin-1/2 spaces.  The total angular momentum operator is the sum
\bea
J^i=W^i+S^i,
\eea
where $W^i$ are spin-1 operators (with multiplication by the unit operator on the spin-1/2 space implied) and $S^i$ are the spin-1/2 operators (with multiplication by the unit operator on the spin-1 space implied).  The projection operators are constructed from the operator
\bea
2 \,{\bf W}\cdot {\bf S}={\bf J}^2-{\bf W}^2-{\bf S}^2=\begin{cases} \phantom{-}1& J=\frac{3}{2}\cr -2 &  J=\frac{1}{2}\, . \cr \end{cases}
\eea
Hence the projection operators ${\cal P}_J$ are
\bea
&&{\cal P}_{1/2}=\sfrac13\left(1-2 \,{\bf W}\cdot {\bf S}\right)\nonumber\\
&&{\cal P}_{3/2}=\sfrac13\left(2+2 \,{\bf W}\cdot {\bf S}\right)\, .
\eea
Using $(W_{jk})^i=-i\epsilon_{ijk}$ and $S^i=\sigma^i/2$ we get
\bea
&&({\cal P}_{1/2})_{jk}=\sfrac13(\delta_{jk}+i\epsilon_{ijk}\sigma^i)=\sfrac13\,\sigma_j\sigma_k\nonumber\\
&&({\cal P}_{3/2})_{jk}=\delta_{jk}-\sfrac13\,\sigma_j\sigma_k\, ,
\eea
leading immediately to the relativistic generalizations
\bea
&&({\cal P}_{1/2})^\alpha{}_\beta=
\frac{1}{3} \left(\gamma -\frac{\not\! P_0 P_0}{m_H^2} \right)^\alpha
\left(\gamma -\frac{\not\! P_0 P_0}{m_H^2} \right)_\beta\nonumber\\
&&({\cal P}_{3/2})^{\alpha}{}_{\beta}=
g^{\alpha}{}_{\beta}- \frac{P_0^\alpha P_{0\beta}}{m_H^2}-({\cal P}_{1/2})^\alpha{}_\beta\, .
\eea
Boosting these from $P_0$ to $P$ gives the operators introduces in Eq.\ 
(\ref{eq:23}) above.

References  \cite{Benmerrouche89,Haberzettl98} define two spin 1/2 projection operators.  In addition to ${\cal P}_{1/2}$ (which they call 
${\cal P}_{11}$) they introduce the operator
\be
({\cal P}_{22})^{\alpha}_{\; \beta} =
\frac{P^\alpha P_\beta}{M^2}\, ,
\ee 
is a spin 1/2 projector on the $(0,\frac{1}{2})$ space.  Since we work in the space of operators satisfying the constraint (\ref{eq:C1}), this operator is excluded from our basis.  In our formalism, it is part of the operator $\tilde g$.

We conclude this appendix by proving the useful relation (\ref{eq:51}).  Start by using the properties
\bea
P_+^\mu D_{\mu\nu}=0=D_{\mu\nu} P_-^\nu\, ,
\eea
which follow directly from the definition of $D_{\mu\nu}$ as a sum over polarizations.  Then we prove two intermediate results.  
First, using the notation ${\cal P}^{\mu\alpha}_{1/2}(P_+)$ 
to denote the projection operator with momentum $P_+$ for a state with mass $M$ [and similarly for ${\cal P}^{\beta\nu}_{1/2}(P_-)$], we see that
\bea
{\cal P}^{\mu\alpha}_{1/2}(P_+)D_{\alpha\beta}{\cal P}^{\beta\nu}_{1/2}=\sfrac19\widetilde \gamma^\mu(P_+)\Big[\gamma^\alpha D_{\alpha\beta} \gamma^\beta\Big] \widetilde\gamma^\nu(P_-)\, .\quad
\label{C9}
\eea
If $b=P_+\cdot P_-$, the quantity in square brackets is
\begin{align}
\big[\cdots\big]&=-4+\frac{\slashed{P}_-\slashed{P}_+}{b} +a\left[\slashed{P}_--\frac{b}{M^2}\slashed{P}_+\right]\left[\slashed{P}_+-\frac{b}{m^2}\slashed{P}_-\right]\nonumber\\
&=-4+\frac{\slashed{P}_-\slashed{P}_+}{b} +a\left[\slashed{P}_-+\frac{b}{M}\right]\left[\slashed{P}_++\frac{b}{m}\right]\nonumber\\
&=-4-2ab+\frac{ab^2}{Mm}+\slashed{P}_-\slashed{P}_+\left[\frac1{b} +a\right]\, ,
\end{align}
where we used the fact that $\slashed{P}_-\to -m$ when acting to the right, and $\slashed{P}_+\to -M$ when acting to the left, because both anticommute with the $\widetilde\gamma$ standing to the right and left, and then can be eliminated using the Dirac equation satisfied by the incoming and outgoing states.  Then, using $\slashed{P}_-\slashed{P}_+=2b-\slashed{P}_+\slashed{P}_-\to 2b-Mm$, and the value of $a$ from Eq.\ (\ref{eqAdef}), we get
\bea
\big[\cdots\big]=-3
\eea
and Eq.\ (\ref{C9}) reduces to
\bea
{\cal P}^{\mu\alpha}_{1/2}(P_+)D_{\alpha\beta}P^{\beta\nu}_{1/2}=-\sfrac13\widetilde \gamma^\mu(P_+)\widetilde\gamma^\nu(P_-) \, .
\eea
Similarly, using the same procedure we can show that  
\begin{align}
\widetilde g^{\mu\alpha}D_{\alpha\beta} {\cal P}_{1/2}^{\beta\nu}=-\sfrac13\widetilde \gamma^\mu(P_+)\widetilde\gamma^\nu(P_-) \, .
\end{align}
Combining these results gives the result we seek
\bea
{\cal P}^{\mu\alpha}_{3/2}(P_+)D_{\alpha\beta}
{\cal P}^{\beta\nu}_{1/2}&=&\widetilde g^{\mu\alpha}D_{\alpha\beta} {\cal P}_{1/2}^{\beta\nu}
\nonumber\\
&&-{\cal P}^{\mu\alpha}_{1/2}(P_+)D_{\alpha\beta}
{\cal P}^{\beta\nu}_{1/2}=0\, .\qquad
\eea


\section{An example of the importance of the collinearity condition }
\label{app:D}

To give some insight into the importance of the collinear frame in the definitions of fixed-axis polarization states, consider  a simple example where the initial and final baryon are identical (both have mass $m$) and the four-momenta are not collinear
\ba
P_\pm^\prime =  (E^\prime,p \sin \theta,0,\pm p \cos \theta) \, , \label{D1}
\ea
with $E^\prime=\sqrt{m^2 + {p^2}}$.  We can transform these momenta to a collinear frame by boosting in the $x$ direction using the transformation
\ba
B_x=\left[ \begin{array}{cccc} \cosh \eta & \sinh \eta & 0 &  0  \cr
\sinh\eta & \cosh\eta & 0 & 0 \cr
0 & 0 & 1 & 0 \cr
0 & 0 & 0 & 1\end{array} \right]\, .
\ea
Collinearity is achieved if 
\bea
\sinh \eta\, E' +\cosh\eta \,p\sin\theta=0\, , \label{D3}
\eea
and the resulting collinear four momenta are
\bea
P_\pm=B_x P'_\pm =(E,0,0,\pm p\cos\theta)
\eea
with $E=\sqrt{m^2+p^2\cos^2\theta}$.  

In this example, consider the longitudinal polarization vectors only.  In the collinear frame they are 
\bea
\varepsilon_{_{0 P_\pm}}=\frac{1}{m}(\pm p\cos\theta,0,0,E)\, .
\eea
Their scalar product is
\bea
m^2\varepsilon_{_{0 P_+}}\cdot\varepsilon_{_{0 P_-}}=-p^2\cos^2\theta-E^2=-P_+\cdot P_-\, .
\eea
In the original, non-collinear frame, the longitudinal polarizations  are
\bea
\varepsilon_{_{0 P'_\pm}}&=&B^{-1}_x\varepsilon_{_{0 P_\pm}}
\nonumber\\
&=&\frac{1}{m}(\pm\cosh\eta\; p\cos\theta,\mp\sinh\eta\; p\cos\theta,0,E)\, .\qquad \label{D7}
\eea
Hence, 
\bea
m^2 \varepsilon_{_{0 P'_+}}\cdot\varepsilon_{_{0 P'_-}}&=&-(p^2\,\cos^2\theta+E^2)\nonumber\\
&=&-(2p^2\,\cos^2\theta+m^2)=-P'_+\cdot P'_-\qquad
\eea
showing that the scalar product of the two longitudinal polarization vectors $\varepsilon_{_{0 P_+}}\cdot\varepsilon_{_{0 P_-}}$ is invariant and uniquely defined.

Now, suppose we were to define the longitudinal vectors in the original frame.  One way to do this would be to observe that the two momenta (\ref{D1}) can be obtained from the vectors 
\bea
\widetilde P_\pm=(E',0,0,\pm p) \label{D9}
\eea
by rotation about the $y$ axis by an angle $\pm\theta$
\be
R_{\pm\theta}=
\left[ \begin{array}{cccc} 
1  & 0 & 0 &  0  \cr
0  & \cos \theta & 0 & \pm \sin \theta \cr
0  & 0 & 1 & 0 \cr
0  & \mp \sin \theta & 0 & \cos \theta \end{array} \right].
\ee
The longitudinal vectors corresponding to (\ref{D9})  are
\bea
\widetilde \varepsilon_\pm=\frac{1}{m}(\pm p,0,0,E')
\eea
so the new polarization vectors would be
\bea
\widetilde \varepsilon\,'_\pm&=&R_{\pm\theta}\;\widetilde\varepsilon_\pm\nonumber\\
&=&\frac{1}{m}(\pm p, \pm\sin\theta\,E',0, \cos\theta\,E')\, .
\eea
These vectors are completely different than the correct ones given in Eq.~(\ref{D7}), and in particular
\bea
m^2\,\widetilde \varepsilon\,'_+\cdot\widetilde \varepsilon\,'_-&=&-(p^2-\sin^2\theta\,E'^2+\cos^2\theta\,E'^2)\nonumber\\
&=&-(2p^2\cos^2\theta +m^2\cos2\theta)\nonumber\\
&\ne& -P'_+\cdot P'_-
\eea

In conclusion: the correct way to treat fixed-axis polarization vectors is to transform to a collinear frame (if necessary), define the fixed-axis vectors there, and then do the inverse transformation back to the original frame (if desired).


\section{$D^{\mu \nu}$ in the collinear frame}
\label{app:E}

Consider the tensor
\be
D^{\mu\nu}=\sum_\lambda \varepsilon_{\lambda P_+}^\mu 
\varepsilon_{\lambda P_-}^{\nu \ast} 
\label{eq1}
\ee
where both $\varepsilon_{\lambda P_+}$ 
are polarization vectors initially oriented along the $\hat z$ axis 
(in a collinear frame) and satisfy the constraints 
$P_+\cdot \varepsilon_{P_+}=P_-\cdot  \varepsilon_{P_-}=0$ 
with 
\be
P_\pm=
\left(E_\pm,0,0,p_\pm \right)
\ee
and $P_+^2=M^2$, $P_-^2=m^2$.  
$D^{\mu \nu}$ is a sum of direct products of four-vectors, and therefore is
a covariant tensor.  
The explicit form of the $\varepsilon_{{\lambda P_\pm}}$ is 
\ba
& &
\varepsilon_{\lambda P_\pm} =\sfrac1{\sqrt{2}}
\left(0,-\lambda ,-i,0 \right) \mbox{\; if \;} \lambda=\pm \nonumber\\
& & 
\varepsilon_{0 P_-} =
\frac{1}{m}
\left( p_-,0,0,E_- \right)
 \nonumber\\
& & 
\varepsilon_{0 P_+} =
\frac{1}{M}
\left( p_+,0,0,E_+ \right).
\ea

Using these explicit forms and interpreting $\varepsilon_{P_+}^\mu$ 
as a {\it column\/} vector and $\varepsilon_{P_-}^{\nu *}$ 
as a {\it row\/} vector, we get the following matrix form for $D$
\begin{equation}D^{\mu\nu}=
\frac{1}{M m}\left[\begin{array}{cccc}
p_+p_- &0 &0 &p_+E_-\cr
 0 & M m & 0&0\cr
 0&0& M m&0\cr
 E_+p_- &0&0&E_+E_-
\end{array}\right]. \label{eq2}
\end{equation}

The covariant form for this tensor can be found 
by exploiting the fact that  $P_{+\mu}D^{\mu\nu}=0$ 
and $D^{\mu\nu}P_{-\nu}=0$.  
Hence the most general form of $D^{\mu\nu}$ is
\ba
D^{\mu\nu}&=&a_1\left(-g^{\mu\nu} + 
\frac{P_-^\mu P_+^\nu}{b}\right)+ \nonumber \\
&+ &a_2\left(P_--\frac{bP_+}{M^2}\right)^\mu
\left(P_+-\frac{bP_-}{m^2}\right)^\nu
\label{E5}
\ea
where $b=P_+\cdot P_-$ and $a_1=1$ 
(to give the correct $D^{xx}$ and $D^{yy}$ components).  
The coefficient $a_2$ can be found from the trace
$$D^\mu{}_\mu=-2-\frac{P_+\cdot P_-}{M m}$$
which gives
$$a_2=-\frac{M m}{b(M m +b)}.$$
It is easy to verify that the two forms (\ref{eq2}) and (\ref{E5}) are identical.


\section{The Current ${J^\mu}$ for the $\gamma$N$\to \Delta$ transition}
\label{app:F}

Equations (\ref{eq:71}) and (\ref{eq:72}) can be written
\begin{align}
&J_{\Delta N}^\mu (P_+,P_-)\nonumber\\
&\quad
=- \frac{1}{\sqrt{3}} f_v\,\overline w_{\beta} (P_+,\lambda_+)\Big[\gamma^\mu D^{\beta\alpha}\gamma_\alpha\gamma^5\Big] u(P_-,\lambda_-){\cal I}\, ,
\label{F1}
\end{align}
where the form factor $f_v$ was defined in Eq.~(\ref{eqJtil1}) and
\be
{\cal I}= 
\int_k 
\psi_\Delta(P_+,k) \psi_N(P_-,k).
\ee

The operator in the square brackets in Eq.~(\ref{F1}) is reduced using the form of $D^{\beta \alpha}$ given in Eq.\ 
(\ref {eqDMm}), remembering that the properties of the Rarita-Schwinger wave function imply that terms proportional to $P_+^\beta$ and $\gamma^\beta$ (when operating to the left) are zero, and that the Dirac equation may be used to replace $\slashed{P}_-\to m$ when operating to the right, and $\slashed{P}_+\to M$ when operating to the left.  We get
\begin{align}
\gamma^\mu D^{\beta\alpha}\gamma_\alpha\gamma^5&=\gamma^\mu\Big[- \gamma^\beta-Aq^\beta\left(\slashed{P}_+-M\right)\Big]\gamma^5\qquad\qquad\nonumber\\
&=\Big[2A(M\gamma^\mu  -P_+^\mu) q^\beta-2g^{\mu\beta}\Big]\gamma^5 \label{F3}
\end{align}
where 
\be
A=\frac{1}{Mm+b}=\frac{2}{(M+m)^2+Q^2}.
\ee
Noting that $P_+^\mu=P^\mu+\sfrac12 q^\mu$, the operator (\ref{F3}) can be written in terms of the invariants of (\ref{eqJS})
\begin{align}
\gamma^\mu &D^{\beta\alpha}\gamma_\alpha\gamma^5
\nonumber\\
&=\Big[g_1 q^\beta \gamma^\mu 
+g_2 q^\beta P^\mu +g_3 q^\beta q^\mu -g_4 g^{\mu \nu}\Big]\gamma^5 \label{F5}
\end{align}
where
\ba
g_1&=&2MA \nonumber \\
g_2&=&-2A \nonumber \\
g_3&=&-A \nonumber \\
g_4&=&2\, .
\label{eqGSSp}
\ea
Note that
\be
g_4=(M+m)g_1+ \frac{M^2-m^2}{2} g_2 -Q^2 g_3
\ee
as required by current  conservation, Eq.\  (\ref{eqG4}).  
The Jones and Scadron form factors 
are 
\be
G_i= f_v g_i \frac{{\cal I}}{\sqrt{3}}.
\ee

We conclude that the physical form factors are, within the $S$-wave model,
\ba
& &G_M^\ast (Q^2)= \frac{8m}{3\sqrt{3} (M+m)} f_v {\cal I} 
\label{eqGMSSp}\\
& &G_E^\ast (Q^2)= G_C^\ast (Q^2)=0\, .
\ea

\section{Asymptotic $Q^2$ dependence of the invariant body integrals}
\label{app:G}

In this appendix we discuss the asymptotic 
dependence of the ``body'' integrals 
\be
B_H(Q^2)= \int_k \psi_H(P_+,k) \psi_N(P_-,k),
\ee
where $H=N$ or $\Delta$. \cite{nucleon}.
The high $Q^2$ dependence of $B_H(Q^2)$ determines 
the asymptotic behavior of the nucleon and $N\to\Delta$ form factors.
To simplify the discussion we consider the easiest case, 
when the parameters of wave function are $\beta_1=\beta_2=\beta$ for the nucleon, $\alpha_1=\alpha_2=\alpha$ for the $\Delta$, and  we will sometimes use the notation $\beta_H$, where $\beta_N=\beta$ and $\beta_\Delta=\alpha$.

The integral $B$ is covariant and may be evaluated in any frame.  It is convenient to evaluate it in the ``antilab'' frame, where the final hadron is at rest. In this case the momenta are 
$P_H=(m_H,0,0,0)$ and $P_{q_0}=(E_0,0,0,-q_0)$, with 
$E_0=\sqrt{m^2+q_0^2}$ the nucleon energy in the initial state.  The photon four-momentum is then $q=(m_H-E_0,0,0,q_0)$ with
\be
q_0^2=\left(\frac{Q^2+m_H^2+m^2}{2m_H}\right)^2-m^2\to\frac{Q^4}{4m_H^2}
\label{eqQlab}
\ee
as $Q^2\to\infty$.  With these momenta we can write the body integral as
\be
B_H(Q^2)=
\frac{N_0}{(2\pi)^2}
\int_0^{\infty} 
\frac{k^2 d k }{2m_s\,E_s} \psi_H(P_H,k)  I(Q^2),
\label{G3}
\ee
where 
\be
I(Q^2)=
\int_{-1}^1 
\frac{dz}{
\left( \beta-2 +2\frac{E_0}{m}\frac{E_s}{m_s} 
+2\frac{q_0}{m} \frac{k}{m_s} z \right)^2}.
\ee
In this frame only the initial state 
depends of the angular coordinate $z=\cos \theta$.
Introducing the parameter  
$$\eta= \frac{(\beta-2)mm_s+2E_0E_s}{2q_0k} 
$$
gives
\bea
I(Q^2) &=& \frac{m^2m_s^2}{4q_0^2 k^2} \int_{-1}^1 
 \frac{dz}{(z+\eta)^2}= 
\frac{m^2m_s^2}{2q_0^2 k^2 (\eta^2-1)}\nonumber\\
& \to & \frac{m^2}{2q_0^2} \label{G5}
\eea
as $Q^2\to\infty$.    Motivated by the nonrelativistic definition of the wave function at the origin, we  define the following covariant integral
\be
\widetilde \psi_H(0)\equiv \int_0^{\infty} 
\frac{k^2 d k }{2m_s \,E_s} \psi_H(P_H,k)\, .
\label{G6}
\ee 
Our results can now be expressed in terms of the behavior of this integral.

{\bf Case I:} If the integral (\ref{G6}) exists, then the diquark momentum $k$ is localized, and the limit (\ref{G5}) can be taken under the integral, leading to the result
\bea
\lim_{Q^2\to\infty}B_H(Q^2)&=&\frac{N_0}{(2\pi)^2}\widetilde\psi_H(0)\,I(Q^2)
\nonumber\\
&\to&\frac{N_0}{(2\pi)^2}\widetilde\psi_H(0)\frac{2m_H^2m^2}{Q^4}\, .
\eea
We obtain the interesting (and well known) result that, if cases where the value of the relativistic wave function of one of the hadrons is finite at the origin, the asymptotic from factor is determined by the high momentum behavior of the other wave function.  For the models used in this paper this shows that  the $N\to\Delta$ body form factors go like $Q^{-4}$ are large $Q^2$.

{\bf Case II:} If  $\widetilde \psi(0)$ does not exist, the limit (\ref{G5}) cannot be taken and the analysis of the large $Q^2$ behavior depends on the behavior of the full integral.  
In this case we return to (\ref{G3}) and write ($H=N$ now)
\bea
B_N(Q^2)=
\frac{N^2_0}{(8 \pi)^2} {\cal I}(Q^2)
\label{eqCaseII}
\eea
The integral ${\cal I}$ can be evaluated 
in the diquark energy scaled by the diquark mass
$x=E_s/m_s$ considering $k=m_s \sqrt{x^2-1}$.
As result we have for $H=N$ 
\be
{\cal I} (Q^2)=
\int_1^{\infty} 
\frac{g(x)}{D\,(x + \omega )^2} dx,
\label{eqInt}
\ee
where $g(x)=\sqrt{x^2-1}$; 
$\omega= \sfrac{1}{2}(\beta-2)$
 and the denominator $D$ is
%
\ba
D &=&  
x^2+ 2\omega \frac{E_0}{m} x + \frac{q_0^2}{m^2}
+ \omega^2 \nonumber \\
 &\to & x^2 + 2\omega \frac{q_0}{m} x + \frac{q_0^2}{m^2}.
\ea
The last approximation holds at large $q_0$, 
where all constant terms and terms proportional to 
$x/q_0$ can be neglected, because they are always 
smaller than terms proportional to $x^2$, $(q_0/m)^2$, or $x\,q_0/m$.
Also for large $q_0$ we can re-write  (\ref{eqInt}) as
\ba
{\cal I} (Q^2)\simeq 
\frac{m^2}{q_0^2}
\int_1^{\infty} g(x)
\left\{ \frac{1}{(x+ \omega)^2} -
\frac{1}{x^2+2\omega \sfrac{q_0}{m} x + \sfrac{q_0^2}{m^2}}
\right\} dx \nonumber \\
 - 2 \omega 
\frac{m^3}{q_0^3}
\int_1^{\infty} g(x)
\left\{ \frac{1}{x+ \omega} -
\frac{x+ 2 \omega \sfrac{q_0}{m}-\omega}{x^2+
2\omega \sfrac{q_0}{m} x + \sfrac{q_0^2}{m^2}}
\right\}  dx. \nonumber 
\ea
Note that each term inside the brackets 
diverges but the result is convergent.
The above integrals can be performed 
analytically following the usual  techniques.
Considering $x=1/\cos u$ the integrand function 
becomes  an algebraic function of $\cos u$ and $\sin u$ 
that we integrate analytically 
using the Mathematica program.  
Considering only the leading and next leading terms in $q_0/m$
in the the general expressions, one is left with
\be
{\cal I} \to 
\frac{m^2}{q_0^2} 
\left[ \log \left( \frac{2q_0}{m} \right) 
 -{\cal R}(\omega)
\right],
\ee
where 
\ba
& &
{\cal R} (\omega)=
1 + \frac{\omega}{\sqrt{1-\omega^2}} \times \label{eqR}  \\
& &\left[
2 \tan^{-1} 
\left( 
\frac{1-\omega}{\sqrt{1-\omega^2}}
\right)+
\tan^{-1} 
\left( 
\frac{\sqrt{1-\omega^2}}{\omega} \right)
\right]\, . \nonumber 
\ea

The analytical continuation of the (\ref{eqR}) 
for the case $\omega \ge 1$ (or $\beta \ge 2$) 
is obtained considering the 
relation between logarithms 
and arc-tangent $\log \sfrac{1+i\,x}{1-i\,x}= 2i \tan^{-1}(x)$ 
and the 
replacement 
$\sqrt{1-\omega^2} \equiv -i \sqrt{\omega^2-1}$. 

In conclusion, we can write 
\be
B_N(Q^2) \to
\frac{N_0^2}{(4\pi)^2} 
\frac{m^4}{Q^4} 
\log \frac{Q^2}{m^2}\, . 
\ee
This term is independent of $\beta$.
The nonleading terms in $Q^{-4}$ carry the  dependence 
in the parameter $\beta $ and set the scale of 
the logarithm behavior.

This logarithmic dependence of the nucleon form factors was missed in 
Ref.~\cite{GrossAga}, but this oversight 
does not affect any of the conclusions of that paper.



\vspace*{-0.22in}

\begin{references}

\vspace*{-0.25in}



\bibitem{Jones99}
M.~K.~Jones {\it et al.}  [Jefferson Lab Hall A Collaboration],
Phys.\ Rev.\ Lett.\  {\bf 84}, 1398 (2000).

\bibitem{Gayou01}
O.~Gayou {\it et al.}  [Jefferson Lab Hall A Collaboration],
Phys.\ Rev.\ Lett.\  {\bf 88}, 092301 (2002).


\bibitem{Punjabi05}
  V.~Punjabi {\it et al.},
  Phys.\ Rev.\ C {\bf 71}, 055202 (2005)
  [Erratum-ibid.\ C {\bf 71}, 069902 (2005)]
  [arXiv:nucl-ex/0501018].

\bibitem{Miller06}
  A.~Kvinikhidze and G.~A.~Miller,
  Phys.\ Rev.\ C {\bf 73}, 065203 (2006)
  [arXiv:nucl-th/0603035].
  
\bibitem{nucleon}
  F.~Gross, G.~Ramalho and M.~T.~Pe\~na,
  Phys.\ Rev.\  C {\bf 77}, 015202 (2008).




\bibitem{Hyde-Wright04}
  C.~E.~Hyde-Wright and K.~de Jager,
  Ann.\ Rev.\ Nucl.\ Part.\ Sci.\  {\bf 54}, 217 (2004)
  [arXiv:nucl-ex/0507001].


\bibitem{Arrington06}
  J.~Arrington, C.~D.~Roberts and J.~M.~Zanotti,
  arXiv:nucl-th/0611050.
 
 
\bibitem{MAMI}
  R.~Beck {\it et al.},
  Phys.\ Rev.\ C {\bf 61}, 035204 (2000)
  [arXiv:nucl-ex/9908017];
  T.~Pospischil {\it et al.},
  Phys.\ Rev.\ Lett.\  {\bf 86}, 2959 (2001)
  [arXiv:nucl-ex/0010020];
  D.~Elsner {\it et al.},
  Eur.\ Phys.\ J.\ A {\bf 27}, 91 (2006)
  [arXiv:nucl-ex/0507014];
  S.~Stave {\it et al.},
  arXiv:nucl-ex/0604013.


\bibitem{LEGS}
  G.~Blanpied {\it et al.},
  Phys.\ Rev.\ C {\bf 64}, 025203 (2001);
  G.~Blanpied {\it et al.},
  Phys.\ Rev.\ Lett.\ {\bf 79}, 4337 (1997).


\bibitem{Bates}
  C.~Mertz {\it et al.},
  Phys.\ Rev.\ Lett.\  {\bf 86}, 2963 (2001)
  [arXiv:nucl-ex/9902012];
  N.~F.~Sparveris {\it et al.}  [OOPS Collaboration],
  Phys.\ Rev.\ Lett.\  {\bf 94}, 022003 (2005)
  [arXiv:nucl-ex/0408003].


\bibitem{CLAS02}
  V.~V.~Frolov {\it et al.},
  Phys.\ Rev.\ Lett.\  {\bf 82}, 45 (1999)
  [arXiv:hep-ex/9808024].
  K.~Joo {\it et al.}  [CLAS Collaboration],
  Phys.\ Rev.\ Lett.\  {\bf 88}, 122001 (2002)
  [arXiv:hep-ex/0110007].

\bibitem{CLAS06}
  M.~Ungaro {\it et al.}  [CLAS Collaboration],
  Phys.\ Rev.\ Lett.\  {\bf 97}, 112003 (2006)
  [arXiv:hep-ex/0606042].
  
  \bibitem{Pascalutsa06b}
  V.~Pascalutsa, M.~Vanderhaeghen and S.~N.~Yang,
  Phys.\ Rept.\  {\bf 437}, 125 (2007)
  [arXiv:hep-ph/0609004].


\bibitem{comment}
  F.~Gross, G.~Ramalho and M.~T.~Pe\~na,
  arXiv:0708.0995 [nucl-th]. To appear in Phys.~Rev.~C.




\bibitem{Gross69}
F.~Gross,
Phys.\ Rev.\  {\bf 186}, 1448 (1969).

\bibitem{Gross82} 
F.\ Gross, Phys.\ Rev.\ C {\bf 26}, 2226 (1982).


\bibitem{Gross92}
  F.~Gross, J.~W.~Van Orden and K.~Holinde,
  Phys.\ Rev.\ C {\bf 45}, 2094 (1992).


\bibitem{Stadler97a}
A.~Stadler and F.~Gross, 
Phys.\ Rev.\ Lett.\ {\bf 78}, 26 (1997).

\bibitem{Stadler97b} 
A.~Stadler, F.~Gross, and M.~Frank,
Phys.\ Rev.\ C {\bf 56}, 2396 (1997). 


\bibitem{Gross04b}
F.~Gross, A.~Stadler and M.~T.~Pe\~na,
Phys.\ Rev.\ C {\bf 69}, 034007 (2004).


\bibitem{Adam97}
J.~Adam, F.~Gross, C.~Savkli and J.~W.~Van Orden,
function,''
Phys.\ Rev.\ C {\bf 56}, 641 (1997).

\bibitem{Gross87}
  F.~Gross and D.~O.~Riska,
  Phys.\ Rev.\ C {\bf 36}, 1928 (1987).


\bibitem{Savkli01}
  C.~Savkli and F.~Gross,
  Phys.\ Rev.\  C {\bf 63}, 035208 (2001)
  [arXiv:hep-ph/9911319].





\bibitem{Rarita41}
  W.~Rarita and J.~S.~Schwinger,
  Phys.\ Rev.\  {\bf 60}, 61 (1941).

\bibitem{Milford55}
   F.~J.~Milford
   Phys.\ Rev.\ {\bf 98}, 1488 (1955).










\bibitem{Benmerrouche89}
  M.~Benmerrouche, R.~M.~Davidson and N.~C.~Mukhopadhyay,
  Phys.\ Rev.\  C {\bf 39} (1989) 2339.


\bibitem{Haberzettl98}
  H.~Haberzettl,
  arXiv:nucl-th/9812043.



\bibitem{work} G.\ Ramalho,  M.\ T.\ Pe\~na, and F.\ Gross, in preparation.

\bibitem{GrossAga} 
  F.~Gross and P.~Agbakpe,
  Phys.\ Rev.\ C {\bf 73}, 015203 (2006).



\bibitem{Miller07a}
  A.~Kvinikhidze and G.~A.~Miller,
  arXiv:nucl-th/0701017.


\bibitem{JacobWick}
  M.~Jacob and G.~C.~Wick,
  Annals Phys.\  {\bf 7}, 404 (1959)
  [Annals Phys.\  {\bf 281}, 774 (2000)].



\bibitem{Carlson}
  C.~E.~Carlson,
  Phys.\ Rev.\ D {\bf 34}, 2704 (1986); 
  C.~E.~Carlson and N.~C.~Mukhopadhyay,
  Phys.\ Rev.\ Lett.\  {\bf 81}, 2646 (1998)
  [arXiv:hep-ph/9804356]; 
  C.~E.~Carlson,
  Few Body Syst.\ Suppl.\  {\bf 11}, 10 (1999)
  [arXiv:hep-ph/9809595].

\bibitem{Jones73}
  H.~F.~Jones and M.~D.~Scadron,
  Annals Phys.\  {\bf 81}, 1 (1973).





\bibitem{Donoghue75}
  J.~F.~Donoghue, E.~Golowich and B.~R.~Holstein,
  Phys.\ Rev.\  D {\bf 12}, 2875 (1975).

\bibitem{Isgur82}
  N.~Isgur, G.~Karl and R.~Koniuk,
  Phys.\ Rev.\  D {\bf 25}, 2394 (1982).


\bibitem{Warns90}
  M.~Warns, W.~Pfeil and H.~Rollnik,
  Phys.\ Rev.\  D {\bf 42}, 2215 (1990).

\bibitem{Capstick94}
  S.~Capstick and B.~D.~Keister,
  Phys.\ Rev.\  D {\bf 51}, 3598 (1995)
  [arXiv:nucl-th/9411016].


\bibitem{Bijker94}
  R.~Bijker, F.~Iachello and A.~Leviatan,
  Annals Phys.\  {\bf 236}, 69 (1994)
  [arXiv:nucl-th/9402012].


\bibitem{JDiaz04}
  B.~Julia-Diaz, D.~O.~Riska and F.~Coester,
  Phys.\ Rev.\  C {\bf 69}, 035212 (2004)
  [Erratum-ibid.\  C {\bf 75}, 069902 (2007)]
  [arXiv:hep-ph/0312169].

\bibitem{JDiaz05}
  B.~Julia-Diaz and D.~O.~Riska,
  Nucl.\ Phys.\  A {\bf 757}, 441 (2005)
  [arXiv:nucl-th/0411012].




\bibitem{JDiaz07a}
 B.~Julia-Diaz, T.~S.~Lee, T.~Sato and L.~C.~Smith,
  Phys.\ Rev.\  C {\bf 75}, 015205 (2007).




\bibitem{Braun06a}
  V.~M.~Braun, A.~Lenz, G.~Peters and A.~V.~Radyushkin,
  Phys.\ Rev.\  D {\bf 73}, 034020 (2006)
  [arXiv:hep-ph/0510237].






\bibitem{Stoler}
  P.~Stoler,
  Phys.\ Rev.\  D {\bf 65}, 053013 (2002)
  [arXiv:hep-ph/0108257];
  P.~Stoler,
  Phys.\ Rev.\ Lett.\  {\bf 91}, 172303 (2003)
  [arXiv:hep-ph/0210184].


\bibitem{Guidal05}
  M.~Guidal, M.~V.~Polyakov, A.~V.~Radyushkin and M.~Vanderhaeghen,
  Phys.\ Rev.\  D {\bf 72}, 054013 (2005)
  [arXiv:hep-ph/0410251].


\bibitem{Pascalutsa06d}
  V.~Pascalutsa, C.~E.~Carlson and M.~Vanderhaeghen,
  Phys.\ Rev.\ Lett.\  {\bf 96}, 012301 (2006)
  [arXiv:hep-ph/0509055].




\bibitem{Burkert04a}
V.~D.~Burkert and T.~S.~H.~Lee,
  Int.\ J.\ Mod.\ Phys.\ E {\bf 13}, 1035 (2004).




\bibitem{Drechsel06a}
  D.~Drechsel and L.~Tiator,
  AIP Conf.\ Proc.\  {\bf 904}, 129 (2007)
  [arXiv:nucl-th/0610112].




\bibitem{SatoLee}
  T.~Sato and T.~S.~H.~Lee,
  Phys.\ Rev.\  C {\bf 54}, 2660 (1996)
  [arXiv:nucl-th/9606009]; 
  T.~Sato and T.~S.~H.~Lee,
  Phys.\ Rev.\  C {\bf 63}, 055201 (2001)
  [arXiv:nucl-th/0010025].


\bibitem{Kamalov}
  S.~S.~Kamalov, S.~N.~Yang, D.~Drechsel, O.~Hanstein and L.~Tiator,
  Phys.\ Rev.\  C {\bf 64}, 032201 (2001)
  [arXiv:nucl-th/0006068];
  S.~S.~Kamalov and S.~N.~Yang,
  Phys.\ Rev.\ Lett.\  {\bf 83}, 4494 (1999)
  [arXiv:nucl-th/9904072].



\bibitem{Pascalutsa}
  G.~L.~Caia, V.~Pascalutsa, J.~A.~Tjon and L.~E.~Wright,
  Phys.\ Rev.\  C {\bf 70}, 032201 (2004)
  [arXiv:nucl-th/0407069];
  V.~Pascalutsa and J.~A.~Tjon,
  Phys.\ Rev.\  C {\bf 70}, 035209 (2004)
  [arXiv:nucl-th/0407068].



\bibitem{Burkert05}
  V.~D.~Burkert,
  Prog.\ Part.\ Nucl.\ Phys.\  {\bf 55} (2005) 108.


\bibitem{JDiaz07b}
 B.~Julia-Diaz, private comunication (2007).






\bibitem{Bartel68}
  W.~Bartel {\it et al.},
  Phys.\ Lett.\ B {\bf 28}, 148 (1968).


\bibitem{Stein75}
  S.~Stein {\it et al.},
  Phys.\ Rev.\ D {\bf 12}, 1884 (1975).








\bibitem{Pascalutsa05}
  V.~Pascalutsa and M.~Vanderhaeghen,
  Phys.\ Rev.\ Lett.\  {\bf 95}, 232001 (2005)
  [arXiv:hep-ph/0508060].


\bibitem{Pascalutsa06c}
  V.~Pascalutsa and M.~Vanderhaeghen,
  Phys.\ Rev.\  D {\bf 73}, 034003 (2006)
  [arXiv:hep-ph/0512244].



\bibitem{Faessler06}
  A.~Faessler, T.~Gutsche, B.~R.~Holstein, V.~E.~Lyubovitskij, D.~Nicmorus and K.~Pumsa-ard,
  arXiv:hep-ph/0612246.

\bibitem{Ramirez07}
  C.~Fernandez-Ramirez, E.~M.~de Guerra and J.~M.~Udias,
  Eur.\ Phys.\ J.\  A {\bf 31}, 572 (2007)
  [arXiv:nucl-th/0611062].

\bibitem{Gail06}
  T.~A.~Gail and T.~R.~Hemmert,
  Eur.\ Phys.\ J.\  A {\bf 28}, 91 (2006)
  [arXiv:nucl-th/0512082].





\bibitem{Alexandrou04}
  C.~Alexandrou, Ph.~de Forcrand, H.~Neff, J.~W.~Negele, 
W.~Schroers and A.~Tsapalis,
  Phys.\ Rev.\ Lett.\  {\bf 94}, 021601 (2005)
  [arXiv:hep-lat/0409122].


\bibitem{Alexandrou07b}
  C.~Alexandrou, G.~Koutsou, H.~Neff, J.~W.~Negele, W.~Schroers and A.~Tsapalis,
  arXiv:0710.4621 [hep-lat].


\end{references}
\end{document}